\documentclass[aps,preprintnumbers,onecolumn,amsmath,amssymb,floatfix,pra]{revtex4-1}

\pdfoutput=1
\usepackage[utf8]{inputenc}

\usepackage{amsmath}
\usepackage{amsfonts}
\usepackage{braket}
\usepackage{tensor}
\usepackage{graphicx}
\usepackage[usenames,dvipsnames,svgnames,table]{xcolor}

\usepackage[colorlinks=true,linkcolor=blue, citecolor=blue, bookmarks]{hyperref}

\def\be{\begin{equation}}
\def\ee{\end{equation}}
\def\bea{\begin{eqnarray}}
\def\eea{\end{eqnarray}}

\begin{document}

\title{Exact solution for the quench dynamics of a nested integrable system}
\author{M\'arton Mesty\'an}
\author{Bruno Bertini}
\author{Lorenzo Piroli}
\author{Pasquale Calabrese}
\affiliation{SISSA and INFN, via Bonomea 265, 34136 Trieste, Italy. }

\begin{abstract}
Integrable models provide an exact description for a wide variety of physical phenomena. 
For example nested integrable systems contain different species of interacting particles with a rich phenomenology 
in their collective behavior, which is the origin of the unconventional phenomenon of spin-charge separation. 
So far, however, most of the theoretical work in the study of non-equilibrium dynamics of integrable systems has been focusing on models with an elementary (\emph{i.e.} not nested) Bethe ansatz. In this work we explicitly investigate quantum quenches in nested integrable systems, by generalizing the application of the Quench Action approach. 
Specifically, we consider the spin-$1$ Lai-Sutherland model, described, in the thermodynamic limit, by the theory of two different species of  Bethe-ansatz particles, each one forming an infinite number of bound states. 
We focus on the situation where the quench dynamics starts from 
a simple matrix product state for which the overlaps with the eigenstates of the Hamiltonian are known. 
We fully characterize the post-quench steady state and perform several consistency checks for the validity of our results. 
Finally, we provide predictions for the propagation of entanglement and mutual information after the quench, 
which can be used as signature of the quasi-particle content of the model. 
\end{abstract}

\maketitle

\section{Introduction}\label{sec:intro}

Integrable models provide a simplified conceptual framework where the overwhelmingly complex collective behavior of  
many-body quantum systems can be investigated. 
Although special by definition, they provide an exact characterization for a wide variety of phenomena. 
During the last decade, the interest in integrable models has been reinvigorated by the experimental possibility of engineering 
quantum many-body systems with ultracold atoms \cite{bdz-08,ccgo-11,gbl-13,PolkonikovRMP11} with the 
unique feature of experimental control over  interaction parameters, dimensionality, and isolation. 
These experiments opened the way to the realization of several 
one-dimensional systems which are described, to a very good approximation, by integrable 
models \cite{HLFS07,kww-06,gklk-12,fse-13,lgkr-13,glms-14,langen-15}. %
%
At the same time, on the theoretical side, the study of integrable models out of 
equilibrium \cite{PolkonikovRMP11, GE15,cem-16} has brought, among other results, to a better understanding 
of relaxation mechanisms \cite{cem-16,ef-16,VR:review}, of transport dynamics \cite{BD:review,BCDF:transport, CADY:hydro,  VM:review}
and of entanglement propagation~\cite{cc-05,CC:review, ac-16,kauf,d-17}. 
Concerning the long time dynamics instead, while generic systems locally relax to thermal states~\cite{D91,S94,R08,RS12}, 
integrable systems admit a larger class of possible stationary states, described by the 
\emph{generalized Gibbs ensemble} (GGE)~\cite{VR:review, RDYO07,ef-16}.
A non-trivial transition between these two ensembles has been observed in the late-time dynamics of weakly non-integrable models~\cite{MK:prethermalization, RoschPRL08, KollarPRB11, worm13, MarcuzziPRL13, EsslerPRB14, fagotti-14, konik14, BF15, CTGM:pret, knap15, SmacchiaPRB15, FC15, BEGR:PRL, MenegozJStatMech15, BEGR:long, NIC14,pret,af-17,mmgs-16,cgdm-17,fnr-17}:
this phenomenon has been named \emph{prethermalization} \cite{bbw-04}.

One of the most important conceptual and technical advances for the description of the many-body dynamics of  interacting integrable 
models has been the introduction of the Quench Action approach \cite{caux-16,ce-13}. 
Its main idea is that at long times the local properties of the system can be captured, in the thermodynamic limit, by a representative 
eigenstate of the Hamiltonian. 
In this  formalism, the time evolution of local observables is written as a sum of contributions coming from excitations over the representative eigenstate. 
While the ideas upon which the method is built are simple, its technical implementation to get quantitative predictions might 
be very hard depending both on the initial state and the model under investigation.

The introduction of the Quench Action approach comes as the result of a long and intense theoretical research activity 
in the field of non-equilibrium dynamics. 
At an early age of the field, analytical calculations were mostly performed in free theories \cite{Caza06,BaSc08,CDEO08,scc-09,CaEF11,CoSC13,fe-13b,KoCC14,sc-14,se-12,grd-10,mc-12}, where the technical obstacles are less severe. 
While some  results on interacting models are known by other 
means \cite{cc-06,fcc-09,cg-11,fm-10,pozsgay-11,ia-12, mussardo-13,la-14,d-14,alba-2015,cubero-16}, it is only with the
Quench Action that they could be systematically obtained in many integrable models \cite{BeSE14,BePC16}, 
with a great amount of work focusing on the prototypical examples of the XXZ Heisenberg 
chain \cite{pozsgay-13,wdbf-14,PMWK14,ac-16qa,ppv-17} and the Lieb-Liniger Bose gas \cite{dwbc-14,dc-14,pce-16,NaPC15,Bucc16,npg-17}. 
These models represent the simplest interacting integrable systems that can be solved by Bethe ansatz. 
They display richer physics than free models: it was only thanks to the quench action solutions of the XXZ spin-chain \cite{wdbf-14,PMWK14} that it has been 
possible to discover that the GGE built with known (ultra)local charges \cite{POZS2-13,fe-13,fcec-14} is not describing 
correctly \cite{POZS2-14,ga-14} the steady state 
and that new families of quasi-local conservation laws should be included in the GGE  
\cite{iqdb-16,IDWC15,iqc-16,PoVW17}. 
In turn, finding a complete set of charges became a parallel research field 
\cite{prosen-11,ip-12,prosen-14,zmp-16,imp-15,ppsa-14,fagotti-16,emp-15,impz-16,
dlsb-15,doyon-15,bs-16,vecu-16,cardy-16,pv-16} 
also because when this is known, the stationary state can be built circumventing the quench action solution, as \emph{e.g.} done in \cite{pvc-16,PVCR17}.

As we are now entering in a more mature age of the field, it is natural to wonder whether it is actually possible to extend the analytical 
techniques developed so far to investigate more complex systems and in particular those which can be solved by nested Bethe ansatz. 
In contrast to the XXZ chain or the Lieb-Liniger gas, these models contain multiple species of interacting particles.
The class of nested systems includes the celebrated Hubbard model \cite{efgk-05} and multi-species point-wise interacting 
Fermi and Bose gases \cite{yang-67}. 
The latter are of particular relevance for cold-atomic experimental realizations \cite{pmcl-14} 
and allow us to study spectacular effects such as, \emph{e.g.}, spin-charge separation. 

In this paper we study quantum quenches to a prototypical nested model and explicitly provide non-trivial predictions for entanglement 
propagation after the quench. 
We consider  the spin-$1$ Lai-Sutherland model \cite{lai-74,sutherland-75}, whose Bethe ansatz solution displays  two distinct species 
of particles, each one forming an infinite family of bound states. 
We focus on a simple initial matrix product state which allows us to generalize the application of the Quench Action approach to nested models 
and deal with the corresponding technical difficulties. 
The overlaps between this matrix product state and the Bethe eigenstates have been reported in \cite{dkm-16}, but in a completely different 
context.  
We fully characterize the post-quench steady state and perform several checks for the validity of our results. 
As a numerically measurable prediction, we provide explicit results for the time evolution of entanglement entropy and for the mutual information 
exploiting a conjecture for entanglement dynamics~\cite{ac-16} which so far has been carefully tested only for the XXZ spin chains. 
The entanglement dynamics turn out to provide a signature of different species of particles and can be used for quench spectroscopy 
similarly to what proposed in \cite{kctc-17,rmc-16}. 

The organization of this paper is as follows. 
In Section~\ref{Sec:Model} we introduce the Lai-Sutherland model and review its nested Bethe ansatz solution. 
The initial state is presented in Section~\ref{Sec:Protocol}, together with a brief introduction to the Quench Action approach, while 
Section~\ref{Sec:SteadyState} is devoted to the analysis and solution of the equations characterizing the post-quench steady state. 
In Section~\ref{Sec:entanglement} we investigate the propagation of entanglement entropy after the quench. 
Finally, our conclusions are reported in Section~\ref{Sec:Conclusions}. 
Five appendices contain the most technical aspects of our calculations.

\section{The model and the Bethe ansatz solution}\label{Sec:Model}

\subsection{The Hamiltonian}

We consider the spin-$1$ Lai-Sutherland model \cite{lai-74,sutherland-75}, described by the Hamiltonian  
\be
H_L=\sum_{j=1}^{L}\left[{\bf s}_j\cdot {\bf s}_{j+1}+\left({\bf s}_j\cdot {\bf s}_{j+1}\right)^2\right]-2L\,,
\label{eq:hamiltonian}
\ee
which acts on the Hilbert space $\mathcal{H}_L=h_1\otimes \ldots\otimes h_L$. Here $h_j\simeq \mathbb{C}^3$ is the local (physical) Hilbert space associated with site $j$. The spin-$1$ operators $s^{a}_j$ are given by the standard three-dimensional representation of the $SU(2)$ generators, explicitly
\be
s^x=\frac{1}{\sqrt{2}}\left(\begin{array}{c c c}0&1&0\\1&0&1\\0&1&0\end{array}\right),\quad s^y=\frac{1}{\sqrt{2}}\left(\begin{array}{c c c}0&-i&0\\i&0&-i\\0&i&0\end{array}\right),\quad s^z=\left(\begin{array}{c c c}1&0&0\\0&0&0\\0&0&-1\end{array}\right)\,.
\label{eq:spin_op}
\ee
In the following, we also define the local spin-$1$ basis as
\be
\ket{\Uparrow} =\left(\begin{array}{c}1\\0\\0\end{array}\right),\quad \ket{0} =\left(\begin{array}{c}0\\1\\0\end{array}\right), \quad \ket{\Downarrow} =\left(\begin{array}{c}0\\0\\1\end{array}\right)\,,
\ee
and will use the labeling
\be
|e_1\rangle=\ket{\Uparrow}\,,\quad |e_2\rangle=\ket{0}\,,\quad |e_{3}\rangle=\ket{\Downarrow}\,,
\label{eq:notation_1}
\ee
and
\be
|e_{\alpha_1}e_{\alpha_2}\ldots e_{\alpha_L}\rangle=|e_{\alpha_1}\rangle\otimes |e_{\alpha_2}\rangle\otimes \ldots \otimes |e_{\alpha_L}\rangle\,.
\label{eq:notation_2}
\ee

The Hamiltonian~\eqref{eq:hamiltonian} is expressed in terms of the $SU(2)$ spin-$1$ operators \eqref{eq:spin_op} and is manifestly invariant under action of $SU(2)$. In fact, it is invariant under the action of the larger group $SU(3)$ as it is apparent by considering its algebraic Bethe ansatz construction \cite{kr-81}, as briefly discussed in Appendix~\ref{sec:algebraic_bethe}.

The Lai-Sutherland model should not be confused with the different spin-$1$ integrable chain
\be
H^{\rm B}_L=\sum_{j=1}^{L}\left[{\bf s}_j\cdot {\bf s}_{j+1}-\left({\bf s}_j\cdot {\bf s}_{j+1}\right)^2\right]\,,
\label{eq:babujian}
\ee
which is the Hamiltonian of the $SU(2)$-invariant Babujian-Takhtajan model \cite{takhtajan-82,babujian-82}. 
This model can be analyzed by means of the so called fusion procedure~\cite{krs-81}, starting from the Bethe ansatz solution of the spin-$1/2$ Heisenberg chain; quantum quenches in the model defined by \eqref{eq:babujian} have been considered in \cite{pvc-16}.  To clarify the difference between the two Hamiltonians, it is useful to introduce the operators
\bea
\mathcal{N}_1 &\equiv& \sum_{j=1}^L\Bigl[ (E_2^2)_j + (E_3^3)_j\Bigr]\,, \label{eq:n1}\\
\mathcal{N}_2 &\equiv& \sum_{j=1}^L \Bigl[ (E_3^3)_j\Bigr]\,,\label{eq:n2}
\eea
where we defined 
\be
E^{i}_j\equiv|e_j\rangle\langle e_i|\,.
\ee
When applied to a state, the operator $\mathcal{N}_{1}$ counts the number of spins which are either $\ket{0}$ or $\ket{\Downarrow}$ while $\mathcal{N}_{2}$ counts the number of spins $\ket{\Downarrow}$. It is straightforward to see that these operators are mutually commuting and moreover commute with the Hamiltonian \eqref{eq:hamiltonian}, which follows directly from the $SU(3)$ invariance of $H_L$. On the other hand, these operators \emph{do not} separately commute with the Hamiltonian $H_{L}^{\rm B}$ (\emph{cf}. \eqref{eq:babujian}): only their sum does.

This seemingly innocent difference has drastic consequences on the physics of the two models. While the quasi-particle content and the structure of elementary excitations of the theory defined by \eqref{eq:babujian} are analogous to that of the spin-$1/2$ case, the one of the theory described by \eqref{eq:hamiltonian} is completely different: two different species of elementary excitations emerge.

\subsection{The nested Bethe anstatz solution}\label{sec:solution}

The Hamiltonian \eqref{eq:hamiltonian} is diagonalized by nested Bethe ansatz. Here we briefly sketch the main aspects relevant to our work while we refer to the specialized literature for a systematic treatment \cite{efgk-05, kr-81,johannesson-86,johannesson2-86}.
 
The starting point is to construct a Bethe-state on the chain of length $L$. In the model at hand, this state is parametrized by {\it two} sets of complex parameters called rapidities, ${\boldsymbol k}_N=\{k_j\}_{j=1}^N$ and ${\boldsymbol \lambda}_M=\{\lambda_j\}_{j=1}^{M}$, as follows:
\bea
| \boldsymbol k_N, \boldsymbol \lambda_N\rangle &=& \sum_{1\leq n_1<\ldots <n_N\leq L}\ \sum_{1\leq m_1<\ldots <m_M\leq N}\ 
\sum_{\mathcal P \in \mathcal S^N} \left(\prod_{1\leq r < l \leq N}\frac{k_{\mathcal P (l)}-k_{\mathcal P (r)}-i}{k_{\mathcal P (l)}-k_{\mathcal P (r)}}\right)\nonumber\\
&\times &\braket{\boldsymbol m|\boldsymbol k_{\mathcal P}, \boldsymbol \lambda} \prod_{r=1}^N\left(\frac{k_{\mathcal{P}(r)}+i/2}{k_{\mathcal{P}(r)}-i/2}\right)^{n_r}\prod_{r=1}^{M}(E^{2}_3)_{n_{m_r}}\prod_{s=1}^{N}(E^{1}_2)_{n_{s}}|\Omega\rangle\,.
\label{eq:bethe_state}
\eea
Here we defined the reference state
\be
|\Omega\rangle=|\Uparrow\Uparrow\ldots \Uparrow\rangle\,,
\label{eq:reference_state}
\ee
together with the functions
\bea
\braket{\boldsymbol m|\boldsymbol k_{\mathcal P}, \boldsymbol \lambda} &=& \sum_{\mathcal R \in \mathcal S^M} A(\boldsymbol \lambda_{\mathcal R})\prod_{\ell=1}^M F_{\boldsymbol k_{\mathcal P}}(\lambda_{\mathcal R(\ell)}; m_\ell)\,,\\
F_{\boldsymbol k}(\lambda,s)&=&\frac{-i }{\lambda - k_s - i /2}\prod_{n=1}^{s-1} \frac{\lambda -  k_n + i /2}{\lambda - k_n - i /2}\,,\\
A(\lambda) &=&\prod_{1\leq r < l \leq M} \frac{\lambda_l-\lambda_r- i}{\lambda_l-\lambda_r}\,.
\eea
The numbers of the two kinds of rapidities, $N$ and $M$, must satisfy
\be
L\geq N\,, \qquad N\geq 2M\,,
\ee 
where $L$ is the size of the chain.

In complete analogy with the well-known Bethe ansatz solution of the spin-$1/2$ Heisenberg chain \cite{kbi-93}, one can show that the state \eqref{eq:bethe_state} is an eigenstate of the Hamiltonian \eqref{eq:hamiltonian} provided that the rapidities satisfy a set of non-linear 
quantization conditions, known as nested Bethe equations, which read
\bea
\left(\frac{k_{j}+i/2}{k_{j}-i/2}\right)^{L}&=&\prod_{\substack{p=1 \\ p\neq j}}^{N} \frac{k_j- k_p+ i}{k_j- k_p- i} \prod_{\ell = 1}^{M} \frac{\lambda_\ell- k_j+ i/2}{\lambda_\ell- k_j- i/2} \qquad j=1,\ldots,N\,,\\
1&=&\prod_{j=1}^N \frac{k_j-\lambda_\ell- i/2}{k_j-\lambda_\ell+ i/2}\prod_{\substack{m=1 \\ m\neq \ell }}^M  \frac{\lambda_\ell-\lambda_m- i}{\lambda_\ell-\lambda_m+ i}\,, \qquad \ell=1,\ldots,M\,.
\label{eq:bethe_equations}
\eea
The energy and the momentum of the eigenstate $\ket{{\boldsymbol k}_N,{\boldsymbol \lambda}_M}$ are given by 
\be
E=-\sum_{j=1}^{N}\frac{1}{k_j^2+1/4}\,,\qquad
P=\left[\sum_{j=1}^{N}i\ln\left[\frac{k_{j}+i/2}{k_{j}-i/2}\right]\right]\textrm{mod}\,2\pi\,.\label{eq:energy_eigenvalue}
\ee
Note that since the Hamiltonian \eqref{eq:hamiltonian} is integrable there exist infinitely many local conserved operators (or charges) beyond energy and momentum, which can be constructed by standard techniques \cite{kbi-93}. In analogy with \eqref{eq:energy_eigenvalue}, the expectation value of higher conserved charges on a Bethe state are immediately obtained once the sets of rapidities characterising the state are known. This is discussed in more detail in Appendix~\ref{sec:algebraic_bethe}.

The Bethe states $|\boldsymbol k_N, \boldsymbol \lambda_M\rangle$ are common eigenstates of the Hamiltonian and of the operators $\mathcal{N}_1$ and $\mathcal{N}_2$ introduced in \eqref{eq:n1}, \eqref{eq:n2}. In particular, one has 
\bea
\mathcal{N}_1| \boldsymbol k_N, \boldsymbol \lambda_M\rangle &=& N| \boldsymbol k_N, \boldsymbol \lambda_M\rangle\,,\label{eq:n1eig}\\
\mathcal{N}_2| \boldsymbol k_N, \boldsymbol \lambda_M\rangle &=& M| \boldsymbol k_N, \boldsymbol \lambda_M\rangle\,.\label{eq:n2eig}
\eea

The physical interpretation for the state \eqref{eq:bethe_state} is straightforward in the case where all the rapidities 
$\boldsymbol k_N, \boldsymbol \lambda_M$ are real. In this case $\boldsymbol k_N$ and $\boldsymbol \lambda_M$ can be thought as the rapidities 
of two different species of quasi-particles created on a vacuum represented by the reference state \eqref{eq:reference_state}; 
we will call these two species of quasi-particles ``bare quasi-particles".  
The state \eqref{eq:bethe_state} is then nothing but a scattering state of bare quasi-particles~\cite{ZZ79}. 
Bare quasi-particles of the first species contribute to the energy and momentum of the state, while those of the second species do not (\emph{cf}. Eqs. \eqref{eq:energy_eigenvalue}). Note that the two species of bare quasi-particles do not directly correspond to the two spin-flips $\ket{0}$ and $\ket{\Downarrow}$. 
Pictorially one could imagine that $\ket{0}$ is a bare quasi-particle of the first species and $\ket{\Downarrow}$ splits 
into a bare quasi-particle of the first and one of the second species.

\subsection{String hypothesis and thermodynamic description}

The thermodynamics of integrable models is naturally described within the well-known thermodynamic Bethe ansatz formalism \cite{takahashi-99}. Within this framework, the quasi-particle content of the model emerges in analogy with the case of non-interacting spin chains.

The thermodynamic Bethe ansatz for the nested system of interest in this work has been widely studied in the literature, see, \emph{e.g.}, Refs.~\cite{johannesson-86,johannesson2-86, afl-83,jls-89,mntt-93,dn-98}. Here we review the aspects which are relevant for our work. As usual, the starting point is provided by the so called string hypothesis, according to which, for large $L$, both sets of rapidities arrange themselves in the complex plane forming patterns called strings. In the present case the parametrization of the strings reads
\bea
k^{n,\ell}_{\alpha}&=&k^{n}_{\alpha}+i\left(\frac{n+1}{2}-\ell\right)\,,\quad \ell=1,\ldots\, n\,,\qquad \alpha=1,\ldots,M_{n}^{(1)}\,,\label{eq:string_1}\\
\lambda^{n,\ell}_{\alpha}&=&\lambda^{n}_{\alpha}+i\left(\frac{n+1}{2}-\ell\right)\,,\quad \ell=1,\ldots\, n\,,\qquad \alpha=1,\ldots,M_{n}^{(2)}\,.\label{eq:string_2}
\eea
Here the numbers $n=1,2,\ldots, +\infty$ are labeling the string types, the real numbers $k^{n}_{\alpha}$, $\lambda^{n}_{\alpha}$ are the string centers and $\{M^{(1)}_n, M^{(2)}_n\}$ are respectively the number of strings of the first and of the second species.  Note that the strings here do \emph{not}
couple the two different species of rapidities and for each species the strings have the the well-known structure encountered in the case of $XXX$ spin-$1/2$ chain \cite{takahashi-99}. Physically, the different strings are interpreted as bound states of the bare quasi-particles (see \emph{e.g.} Chapter IV in Ref.~\cite{efgk-05}).

Under the string hypothesis, the Bethe equations \eqref{eq:bethe_equations} can be turned into equations for the string centers; it is convenient to consider the logarithmic form of these equations which reads as~\cite{johannesson2-86}
\begin{align}
&z^{(1)}_{n}(k_\alpha^{n})=\frac{2 \pi}{L} I^{n}_{\alpha}\,,& &z^{(2)}_{n}(\lambda_\alpha^{n})=\frac{2 \pi}{L} J^n_{\alpha}\,.
\label{eq:logarithmicBTE}
\end{align}
Here $\{I^n_{\alpha}\},\{J^n_{\alpha}\}$ are integers or semi-integers depending on $\{M^{(1)}_n, M^{(2)}_n\}$ and we introduced the \emph{counting functions} 
\begin{align}
&z^{(1)}_{n}(\lambda)=p_n(\lambda)-\frac{1}{L}\sum_{m=1}^{+\infty}\sum_{\beta=1}^{M^{(1)}_m}\Xi_{nm}(\lambda-k^{m}_{\beta})+\frac{1}{L}\sum_{m=1}^{+\infty}\sum_{\beta=1}^{M^{(2)}_m} \Theta_{nm}(\lambda-\lambda^{m}_{\beta})\,, \label{eq:counting1}\\
&z^{(2)}_{n}(\lambda)=\frac{1}{L}\sum_{m=1}^{+\infty}\sum_{\beta=1}^{M^{(1)}_m}\Theta_{nm}(\lambda-k^{m}_{\beta})-\frac{1}{L}\sum_{m=1}^{+\infty}\sum_{\beta=1}^{M^{(2)}_m} \Xi_{nm}(\lambda-\lambda^{m}_{\beta})\,.\label{eq:counting2}
\end{align}
Here
\bea
   p_n(\lambda)&=&2\arctan\left(\frac{2\lambda}{n}\right)\,, \label{eq:pn}\\
\Xi_{n,m}(\lambda)&=&(1-\delta_{nm})p_{|n-m|}(\lambda)+2p_{|n-m|+2}(\lambda)+\ldots+2p_{n+m-2}(\lambda)+p_{n+m}(\lambda)\,,\label{eq:Xi_mn}\\
\Theta_{n,m}(\lambda)&=&p_{|n-m|+1}(\lambda)+p_{|n-m|+3}(\lambda)+\ldots+p_{n+m-1}(\lambda)\,.
\label{eq:Theta_mn}
\eea
The counting functions \eqref{eq:counting1} and \eqref{eq:counting2} are assumed to be monotonic for any solution of the equations \eqref{eq:logarithmicBTE}. The range of $\{I_\alpha^n,J_\beta^m\}$ can be obtained taking the $\lambda\rightarrow\infty$ limit of the counting functions (and imposing $\{I_\alpha^n,J_\beta^m\}$ to be integers or semi-integers according to the values of $\{M^{(1)}_n, M^{(2)}_n\}$). The result reads as~\cite{dn-98} 
\bea
|I_\alpha^n| &\leq&\frac{1}{4}\left(2L-2+M^{(2)}_n-2\sum_{m=1}^{+\infty}t_{nm}{M^{(1)}_m}+\sum_{m=1}^{+\infty} t_{nm}{M^{(2)}_m} \right)\,,\\
|J_\alpha^n| &\leq&\frac{1}{4}\left(M^{(1)}_{n}-2+\sum_{m=1}^{+\infty}t_{nm}{M^{(1)}_m}-2\sum_{m=1}^{+\infty} t_{nm}{M^{(2)}_m} \right)\,,
\eea
where $t_{nm}\equiv2\min(n,m)-\delta_{n,m}$. The energy and the momentum of a state described by the string centers $\{k_\alpha^n\},\{\lambda^n_\alpha\}$ are given by
\be
E =\sum_{m=1}^{+\infty}\sum_{\beta=1}^{M^{(1)}_m}\varepsilon_m(k_\beta^m)\,,\qquad\qquad
P
 =\left[\sum_{m=1}^{+\infty}\sum_{\beta=1}^{M^{(1)}_m}\left(p_m(k_\beta^m)-\pi\right)\right]\,\text{mod}\,2\pi\,, 
\label{eq:stringenergyandmomentum}
\ee
where we introduced the bare energies~\cite{johannesson-86}
\be
\varepsilon_{n}(k)=-\frac{n}{k^2+n^2/4}\,.
\label{epsilon_energy}
\ee
Finally, the eigenvalues of $\mathcal N_1$ and $\mathcal N_2$ read as 
\be
N_1 = \sum_{n=1}^{+\infty}n M^{(1)}_n \,,\qquad\qquad
N_2 = \sum_{n=1}^{+\infty}n M^{(2)}_n\,.
\ee
In the thermodynamic limit the string centers become continuous variables on the real line. Accordingly, we introduce the rapidity distribution functions $\rho^{(1)}_n(k)$ and $\rho^{(2)}_n(k)$ which generalize the momentum distribution functions for free systems. Analogously, we also introduce the hole distribution functions $ \rho^{(1)}_{h,n}(k)$ and $ \rho^{(2)}_{h,n}(k)$, which corresponds to the distribution of holes (\emph{i.e.} values of the rapidity for which there is no particle) in free Fermi gases at finite temperature. As customary, we define
\bea
\rho^{(r)}_{t,n}(k)=\rho^{(r)}_{n}(k)+\rho^{(r)}_{h,n}(k)\,,\quad r=1,2\,,\quad n=1,\ldots , +\infty\,,
\eea
as well as the functions
\bea
\vartheta_n^{(r)}&= &\frac{\rho^{(r)}_{n}(x)}{\rho^{(r)}_{t,n}(x)}\,,\quad r=1,2\,,\quad n=1,2,\ldots ,+\infty\,,\\
\eta^{(r)}_n(x) & = &\frac{\rho^{(r)}_{h,n}(x)}{\rho^{(r)}_{n}(x)}\,,\quad r=1,2\,,\quad n=1,2,\ldots ,+\infty\,.
\label{eq:eta_functions}
\eea
Particle and hole distribution functions of the two species can not be chosen arbitrarily, but are constrained by a set of linear integral equations which are nothing but the thermodynamic version of the Bethe equations \eqref{eq:bethe_equations}. They are usually called Bethe-Takahashi equations and are derived in complete analogy with the XXX spin-$1/2$ case \cite{takahashi-99}. They read \cite{mntt-93}
\bea
\rho_{t,n}^{(1)}(\lambda)&=&a_n(\lambda)-\sum_{m=1}^{\infty}\left(a_{n,m}\ast\rho_m^{(1)}\right)(\lambda)+\sum_{m=1}^{\infty}\left(b_{n,m}\ast\rho_m^{(2)}\right)(\lambda)\,,\label{eq:TBAexplicit1}\\
\rho_{t,n}^{(2)}(\lambda)&=&-\sum_{m=1}^{\infty}\left(a_{n,m}\ast\rho_m^{(2)}\right)(\lambda)+\sum_{m=1}^{\infty}\left(b_{n,m}\ast\rho_m^{(1)}\right)(\lambda)\,.\label{eq:TBAexplicit2}
\eea
Here we defined the convolution between two functions as
\be
\left(f\ast g\right)(\lambda)=\int_{-\infty}^{\infty}{\rm d}\mu f(\lambda-\mu)g(\mu)\,,
\label{eq:convolution}
\ee
and
\bea
a_{n,m}(\lambda)&=&(1-\delta_{nm})a_{|n-m|}(\lambda)+2a_{|n-m|+2}(\lambda)+\ldots+2a_{n+m-2}(\lambda)+a_{n+m}(\lambda)\,,\label{eq:a_mn}\\
b_{n,m}(\lambda)&=&a_{|n-m|+1}(\lambda)+a_{|n-m|+3}(\lambda)+\ldots+a_{n+m-1}(\lambda)\,,
\label{eq:b_mn}
\eea
where
\bea
a_{n}(\lambda)&=&\frac{1}{2\pi}\frac{n}{\lambda^2+n^2/4}\,.
\label{eq:a_function}
\eea
Following \cite{takahashi-99}, the Bethe-Takahashi equations \eqref{eq:TBAexplicit1}, \eqref{eq:TBAexplicit2} can also be cast in partially decoupled form which is more convenient for numerical analysis. We refer to Appendix~\ref{sec:general_calculations} for the derivation, while here we only report the final result
\bea
\rho_{t,n}^{(1)}(\lambda)&=&\delta_{n,1}s(\lambda)+s\ast\left( \rho_{h,n-1}^{(1)}+\rho_{h,n+1}^{(1)}\right)(\lambda)+s\ast \rho_{n}^{(2)}(\lambda)\,,\label{eq:tri-diagonal_BT1}\\
\rho_{t,n}^{(2)}(\lambda)&=&s\ast\left( \rho_{h,n-1}^{(2)}+\rho_{h,n+1}^{(2)}\right)(\lambda)+s\ast \rho_{n}^{(1)}(\lambda)\,,
\label{eq:tri-diagonal_BT2}
\eea
which uses the conventions
\be
\rho^{(r)}_{h,0}(\lambda)\equiv 0\,,\qquad\qquad r=1,2\,,
\label{eq:convention_rhot0}
\ee
and the function
\be
s(\lambda)=\frac{1}{2{\cosh}\left(\pi\lambda\right)}\,.
\label{eq:s_function}
\ee
The set of rapidity distribution functions completely characterizes the thermodynamic properties of a given macrostate. In particular, the density of the quasi-particles of the species $(1)$ and $(2)$ can be computed as
\bea
D^{(1)}&=& \frac{{N}_{1}}{L}=\sum_{n=1}^{+\infty}n\int_{-\infty}^{+\infty}\!\!\!\!{\rm d}k\,\, \rho_{n}^{(1)}(k)\,,\label{eq:density1}\\
D^{(2)}&=&\frac{{N}_{2}}{L}=\sum_{n=1}^{+\infty}n\int_{-\infty}^{+\infty}\!\!\!\!{\rm d}\lambda\,\, \rho_{n}^{(2)}(\lambda)\,.\label{eq:density2}
\eea
For later convenience, it is also useful to introduce the density of particles forming bound states as
\be
D^{(r)}_{n}=n\int_{-\infty}^{+\infty}\!\!\!\!{\rm d}k\,\, \rho_{n}^{(r)}(k)\,.
\label{eq:string_content}
\ee
Analogously, by means of the string hypothesis, one can obtain the density of energy from \eqref{eq:stringenergyandmomentum} as
\be
\frac{E}{L}=\sum_{n=1}^{+\infty}\int_{-\infty}^{+\infty}\!\!\!\!{\rm d}k\,\, \rho_{n}^{(1)}(k)\varepsilon_{n}(k)\,.
\ee

\section{The quench protocol}\label{Sec:Protocol}

We are now interested in the standard quench dynamics where the system is prepared in an initial state 
$\ket{\Psi_0}$, which is not an eigenstate of the Hamiltonian, and then it is let evolve for $t>0$ with 
the Lai-Sutherland Hamiltonian \eqref{eq:hamiltonian}. The main goal of our analysis is not to reconstruct the 
entire time evolution, which is currently out of reach, but just to have an exact characterization of the stationary state. 

As we have mentioned in the introduction, to tackle this problem, we have two possible strategies. 
Either we rely on the knowledge of the expectation value of a complete set of quasi-local charges or we 
construct the stationary state from the Quench Action. 
Concerning the former approach,  while we expect that quasi-local charges might be successfully employed also in the study of quenches to 
the $SU(3)$ chain \eqref{eq:hamiltonian}, a systematic analysis  of this problem (\emph{i.e.} a systematic characterization and classification 
of the quasi-local charges) in the case of nested systems has not yet been carried out. 
As a consequence, we can only rely on the Quench Action approach~\cite{ce-13}, which can be implemented quite generally for any 
Bethe ansatz integrable model, but is limited to those quenches for which the overlap between the initial state and the Bethe 
eigenstates are known exactly. 
The computation of the overlaps has in fact turned out to be a very hard technical problem, and no general scheme to solve it  
has yet been developed \cite{kp-12,pozsgay-14,bdwc-14,pc-14,fz-16,msca-16,hst-17,cd-14}. 
In the following we then restrict ourselves to consider a special initial state for which these overlaps are known. 
Indeed, an exact formula has been conjectured and tested in \cite{dkm-16} in the context of the AdS/CFT correspondence. 
While the special initial state that we consider is admittedly artificial, we will see that it has a very simple structure. 
This, in turn, makes it possible to test our theoretical predictions by means of efficient numerical methods. 

Before presenting the specific initial state considered and analysing the corresponding overlap formula, 
let us briefly sketch the main aspects of the Quench Action method which are relevant for our purposes. 

\subsection{The Quench Action approach}

Despite its introduction being still quite recent \cite{ce-13}, the Quench Action approach has already been employed for many models and 
quench settings. It is not our intention here to provide a detailed review of the method, so we simply sketch the key formulas in order to set the notations. A pedagogical introduction can be found in \cite{caux-16}.

In the equilibrium case, at finite temperature, a well-established result of the thermodynamic Bethe ansatz is that the thermodynamic 
properties of an integrable system are encoded in the rapidity distribution functions of a representative eigenstate. 
The latter are determined as the solution of a set of non-linear integral equations which are derived by minimizing the thermal free 
energy functional \cite{takahashi-99}. 
Remarkably, Ref.~\cite{ce-13} showed that the same idea can be applied to study the stationary state describing local observables 
after a quantum quench. 
In this case, the thermal free energy is replaced by another functional called Quench Action.  
In complete analogy to the thermal case, the saddle-point rapidity distributions fully characterize the thermodynamic properties of the 
post-quench steady state. Relevant examples of such properties are the local correlations and the velocities of the quasi-particle excitations.

The Quench Action functional explicitly depends on the initial state $|\Psi_0\rangle$ and on some symmetry properties of the latter. 
In the case considered in this work it reads 
\be 
S_{\rm QA}[{\boldsymbol{\rho}}]=2S_{\Psi_0}[{\boldsymbol{\rho}}]-\frac{1}{2}S_{\rm YY}[{\boldsymbol{\rho}}]\,,
\label{eq:SQA}
\ee
where we have indicated with compact notation ${\boldsymbol \rho}$ the sets $\{\rho^{(1)}_n\}_{n=1}^{\infty}$, $\{\rho^{(2)}_n\}_{n=1}^{\infty}$. Here $S_{\rm YY}[{\boldsymbol\rho}]$ is the so called Yang-Yang entropy \cite{johannesson-86}
\be
S_{\rm YY}\left[{\boldsymbol{\rho}}\right]=\sum_{r=1}^{2}\sum_{n=1}^{+\infty}\int_{-\infty}^{+\infty}{\rm d}x \left\{\left(\rho_n^{(r)}(x)+\rho_{h,n}^{(r)}\right)\ln\left(\rho_n^{(r)}(x)+\rho_{h,n}^{(r)}\right)-\rho_{n}^{(r)}\ln\rho_{n}^{(r)}-\rho_{h,n}^{(r)}\ln\rho_{h,n}^{(r)}
\right\}\,,
\label{eq:yangyang}
\ee
while we will comment in the next subsection on the factor $1/2$ in front of $S_{\rm YY}[{\boldsymbol \rho}]$ appearing in \eqref{eq:SQA}. The first term in \eqref{eq:SQA} is defined in terms of the thermodynamically leading part of the overlap between the initial state $|\Psi_0\rangle$ and the Bethe state corresponding to $|{\boldsymbol \rho}\rangle$ as follows
\be
S[{\boldsymbol{\rho}}]=-\lim_{\rm th}\frac{1}{L}{\rm Re}\left[\ln\langle \Psi_0|{\boldsymbol{\rho}}\rangle\right]\,.
\label{eq:therm_overlap}
\ee
This expression has to be interpreted as the thermodynamic limit of the logarithm of the overlap between $|\Psi_0\rangle$ and a Bethe state which corresponds to the functions $\{\rho^{(1)}_n\}_{n=1}^{\infty}$ and $\{\rho^{(2)}_n\}_{n=1}^{\infty}$. The post-quench steady state is then described by the distributions $\bar{\rho}^{(r)}_{n}(\lambda)$ which are the saddle-point of the Quench Action \eqref{eq:SQA}, namely the solution of the system of equations
\be
\frac{\partial S_{QA}[{\boldsymbol{\rho}}]}{\partial \rho^{(r)}_n(\lambda)} \Bigg|_{{\boldsymbol{\rho}}={\bar{\boldsymbol{\rho}}}}=0, \qquad\qquad r=1,2\,,\qquad\qquad n\geq 1.
\label{eq:general_saddle_point}
\ee
In the next subsection, we will explicitly write the Quench Action for the initial state studied in this work, and derive the corresponding saddle-point equations.

\subsection{The initial state and the saddle-point equations}

We consider the following initial state for our quench problem 
\bea
\ket{\Psi_0} &=& \frac{1}{\sqrt\mathcal N}\,{\rm tr}_0\left[\prod_{j=1}^L\Big(\sigma^1\ket{\Uparrow}_j+\sigma^2\ket{0}_j+\sigma^3\ket{\Downarrow}_j\Big)\right]
=\frac{1}{\sqrt\mathcal N} \sum_{\{\alpha_j\}}{\rm tr}_0\left[\sigma^{\alpha_1}\sigma^{\alpha_2}\ldots \sigma^{\alpha_L}\right]\ket{e_{\alpha_1}e_{\alpha_2}\ldots e_{\alpha_L}}\,.
\label{eq:initial_state}
\eea
Here we used the notations \eqref{eq:notation_1} and \eqref{eq:notation_2}, while $\sigma^{\alpha}$ are the Pauli matrices
\be
\sigma^{1}=\left(\begin{array}{c c}0&1\\1&0\end{array}\right)\,,\quad \sigma^{2}=\left(\begin{array}{c c}0&-i\\i&0\end{array}\right)\,,\quad \sigma^{3}=\left(\begin{array}{c c}1&0\\0&-1\end{array}\right)\,,
\ee
so that the trace in \eqref{eq:initial_state} is over the auxiliary space $h_0\simeq \mathbb{C}^{2}$. Finally, the normalization $\mathcal N=3^L+3 (-1)^L$ is chosen such that $\braket{\Psi_0|\Psi_0}=1$. 

The initial state \eqref{eq:initial_state} is a matrix product state with local dimension $2$ which satisfies cluster decomposition.
 The simple structure of $\ket{\Psi_0}$ allows for the investigation of its time evolution by means of efficient numerical methods such as DMRG and iTEBD simulations \cite{wf-04,dksv-04,schollwock-05}. 
Consider a chain of length $L$ and a Bethe state characterized by the rapidities $\{k_j\}_{j=1}^{N}, \{\mu_j\}_{j=1}^{M}$ such that $L$, $N$ and $M$ are even numbers and
\bea
\{k_j\}_{j=1}^{N}&=&\{-k_j\}_{j=1}^{N} \label{eq:parity1}\,,\\
\{\mu_j\}_{j=1}^{M}&=&\{-\mu_j\}_{j=1}^{M}\label{eq:parity2}\,.
\eea
Then, its overlap with the initial state \eqref{eq:initial_state} is given by \cite{dkm-16}
\be
\langle\Psi_0|\{k_j\}_{j=1}^{N}, \{\mu_j\}_{j=1}^{M}\rangle=\frac{2}{\sqrt{\mathcal N}}\sqrt{\left[\prod_{m=1}^{N/2}\frac{k_m^2+1/4}{k_m^2}\right]\left[\prod_{m=1}^{M/2}\frac{\lambda_m^2+1/4}{\lambda_m^2}\right]\frac{{\rm det} G_{+}}{{\rm det} G_{-}}}\,.
\label{eq:overlap_formula}
\ee
Here $G_{\pm}$ are Gaudin-like matrices defined by
\bea
G_{\pm}=\left(
\begin{array}{cc}
A_{\pm}& B_\pm\\
B^{t}_{\pm}& C_{\pm}
\end{array}
\right),
\eea
where
\bea
\left(A_{\pm}\right)_{r,s}&=&\delta_{rs}\left[L \mathcal{K}_{1}(k_r)-\sum_{l=1}^{N/2} \mathcal{K}^{+}_{2}(k_r, k_l)+\sum_{l=1}^{M/2} \mathcal{K}_1^{+}(k_r, \lambda_l)\right]+\mathcal{K}^{\pm}_{2}(k_r,k_s)\,,\\
\left(B_{\pm}\right)_{r,s}&=&-\mathcal{K}^{\pm}_{1}(k_r,\lambda_s)\,,\\
\left(C_{\pm}\right)_{r,s}&=&\delta_{rs}\left[-\sum_{l=1}^{M/2} \mathcal{K}^{+}_{2}(\lambda_r, \lambda_l)+\sum_{l=1}^{M/2} \mathcal{K}_1^{+}(\lambda_r, k_l)\right]+\mathcal{K}^{\pm}_{2}(\lambda_r,\lambda_s)\,,
\eea
with the additional definitions
\bea
\mathcal{K}_{1}(u)&=&\frac{1}{u^2+1/4}\,,\\
\mathcal{K}_{2}(u)&=&\frac{2}{u^2+1}\,,\\
\mathcal{K}_{s}^{\pm}(u,w)&=&\mathcal{K}_{s}(u-w)\pm \mathcal{K}_{s}(u+w)\,,\quad s=1,2\,.
\eea
An analogous formula exists for the case where $N$ is even, while $L$ and $M$ are odd, and the sets of rapidity distributions still satisfy \eqref{eq:parity1}, \eqref{eq:parity2}. Conversely, for Bethe states not satisfying \eqref{eq:parity1}, \eqref{eq:parity2} the overlap is zero \cite{dkm-16}. Formula \eqref{eq:overlap_formula} was conjectured in \cite{dkm-16} based on an analogy with the case of the $XXX$ spin-$1/2$ chain, where a similar state can be constructed and the corresponding overlaps computed \cite{LeKZ15,BLKZ16,fz-16}.

It is not simple to extract from \eqref{eq:overlap_formula} the thermodynamically leading part of the overlap. In fact, due to divergences arising in the matrices $G_{\pm}$, one should take into account finite-size deviations from perfect strings \eqref{eq:string_1}, \eqref{eq:string_2}. Note however that the situation is completely analogous to the one encountered in other models displaying bound states. This is, for instance, the case for quenches from the N\'eel state to the $XXZ$ spin-$1/2$ chain \cite{wdbf-14,PMWK14} or from non-interacting to attractive one-dimensional Bose gases \cite{pce-16}. Following these works, it can be argued that the ratio of the determinants in \eqref{eq:overlap_formula} only gives a sub-leading contribution in the thermodynamic limit and can thus be neglected. Given the similarity of the argument, we do not report it here, and refer to \cite{wdbf-14,PMWK14, pce-16} for more details.

Dropping the ratio of the determinants, it is straightforward to take the thermodynamic limit of \eqref{eq:overlap_formula}. Since the calculations are analogous to the ones performed in \cite{wdbf-14, pce-16}, here we only report the final result, which reads
\be
S_{\Psi_0}\left[\{\rho^{(r)}_n\}_{n=1}^{\infty}\right]\equiv -\ln\left(\langle\Psi_0| \{\rho^{(r)}_n\}_{n=1}^{\infty}\rangle \right) =\frac{1}{2}\ln3+\frac{1}{4}\sum_{n=1}^{\infty}\int_{-\infty}^{\infty}{\rm d}k \rho^{(1)}_{n}(k) g_{n}(k)+\frac{1}{4}\sum_{n=1}^{\infty}\int_{-\infty}^{\infty}{\rm d}\lambda \rho^{(2)}_{n}(\lambda)g_{n}(\lambda)\,,
\label{eq:sq_expression}
\ee
where we defined
\bea
g_n(\lambda) &=& \sum_{l=0}^{n-1} \Big[ f_{n-1-2l}(\lambda) - f_{n-2l}(\lambda) \Big]\,, \label{eq:g_function}\\ 
f_n (\lambda) &=& \ln \big(\lambda^2 + n^2/4 \big) \label{eq:f_function}\,.
\eea

From \eqref{eq:SQA}, we see that we have now all the elements necessary to explicitly write down the saddle-point equations \eqref{eq:general_saddle_point}. Note that the factor $1/2$ in front of the Yang-Yang entropy \eqref{eq:yangyang} comes from the fact that the initial state $|\Psi_0\rangle$ has non-vanishing overlaps only with parity-symmetric Bethe states \cite{dwbc-14}. Putting all together, we can explicitly perform the functional derivative in \eqref{eq:general_saddle_point} and obtain the desired saddle-point equations. By exploiting the Bethe-Takahashi equations \eqref{eq:TBAexplicit1} and \eqref{eq:TBAexplicit2} we finally obtain
\bea
\ln \eta_{n}^{(1)}(\lambda)&=&g_n(\lambda)+\sum_{m=1}^{+\infty}\left[a_{n,m}\ast \ln\left(1+\left[\eta_m^{(1)}\right]^{-1}\right)\right](\lambda)-\sum_{m=1}^{+\infty}\left[b_{n,m}\ast \ln\left(1+\left[\eta_m^{(2)}\right]^{-1}\right)\right](\lambda)\,,\label{eq:eta_(1)}\\
\ln \eta_{n}^{(2)}(\lambda)&=&g_n(\lambda)+\sum_{m=1}^{+\infty}\left[a_{n,m}\ast \ln\left(1+\left[\eta_m^{(2)}\right]^{-1}\right)\right](\lambda)-\sum_{m=1}^{+\infty}\left[b_{n,m}\ast \ln\left(1+\left[\eta_m^{(1)}\right]^{-1}\right)\right](\lambda)\,,\label{eq:eta_(2)}
\eea
where the functions $a_{n,m}$ and $b_{n,m}$ are defined in \eqref{eq:a_mn} and \eqref{eq:b_mn} respectively.

Once again, Eqs.~\eqref{eq:eta_(1)} and \eqref{eq:eta_(2)} can be cast in a partially decoupled form which is more suitable for numerical evaluation, \emph{i.e.},
\bea
\ln \eta_n^{(1)}(\lambda)&=&d_n(\lambda)+s\ast\left( \ln [1+\eta_{n-1}^{(1)}]+\ln [1+\eta_{n+1}^{(1)}]\right)(\lambda)-s\ast \ln \left[1+\left(\eta_n^{(2)}\right)^{-1}\right](\lambda)\,,\label{eq:simple_decoupled_1}\\
\ln \eta_n^{(2)}(\lambda)&=&d_n(\lambda)+s\ast\left( \ln [1+\eta_{n-1}^{(2)}]+\ln [1+\eta_{n+1}^{(2)}]\right)(\lambda)-s\ast \ln \left[1+\left(\eta_n^{(1)}\right)^{-1}\right](\lambda)\,,\label{eq:simple_decoupled_2}
\eea
where $s(\lambda)$ is defined in \eqref{eq:s_function}, and we introduced
\be
d_n(\lambda)=(-1)^{n+1}\ln\left[\tanh^2\left(\frac{\pi \lambda}{2}\right)\right]\,,
\label{eq:driving_dn}
\ee
with the convention
\bea
\eta^{r}_0(\lambda) &\equiv & 0\,.
\label{eq:convention_eta0}
\eea
Interestingly, Eqs.~\eqref{eq:eta_(1)} and \eqref{eq:eta_(2)} can be cast in yet another form, which is still partially decoupled. We report it in Appendix~\ref{sec:derivation_decoupled}, together with the derivation of \eqref{eq:simple_decoupled_1} and \eqref{eq:simple_decoupled_2}. 

One can immediately see that \eqref{eq:eta_(1)} are \eqref{eq:eta_(2)} are symmetric under exchanging the particle species $(1)\leftrightarrow (2)$. This observation allows us to simplify the saddle-point equations as follows. Suppose we find a solution $\eta^{(1)}_n(\lambda)=\Theta_n(\lambda)$, $\eta_n^{(2)}(\lambda)=\Xi_n(\lambda)$. If $\Theta_n(\lambda)\neq \Xi_n(\lambda)$, then we find another solution $\eta^{(1)}_n(\lambda)=\Xi_n(\lambda)$, $\eta^{(2)}_n(\lambda)=\Theta_n(\lambda)$. We rule out this possibility by assuming uniqueness of the solution of \eqref{eq:eta_(1)} and \eqref{eq:eta_(2)}. Hence, we conclude that the two sets of functions coincide, namely
\be
\eta^{(1)}_n(\lambda)=\eta^{(2)}_n(\lambda)\equiv \eta_n(\lambda)\,.\label{eq:etas}
\ee
As a consequence, one can write a unique set of non-linear integral equations for $\eta_n(\lambda)$. From \eqref{eq:tri-diagonal_BT1} and \eqref{eq:tri-diagonal_BT2} it follows
\bea
\ln \eta_n(\lambda)&=d_n(\lambda)+s\ast\left( \ln [1+\eta_{n-1}]+\ln [1+\eta_{n+1}]\right)(\lambda)-s\ast \ln \left[1+\eta_n^{-1}\right](\lambda)\,.
\label{eq:final_decoupled}
\eea
The corresponding root densities $\rho^{(1)}_n(\lambda)$ and $\rho^{(2)}_n(\lambda)$ are found by solving the Bethe-Takahashi equations  \eqref{eq:TBAexplicit1}\,--\,\eqref{eq:TBAexplicit2}. Note that, even if $\eta^{(1)}_n(\lambda)=\eta^{(2)}_n(\lambda)$, the root densities will generically be \emph{different} due to the asymmetric form of \eqref{eq:TBAexplicit1}\,--\,\eqref{eq:TBAexplicit2}. 

The single set of decoupled equations \eqref{eq:final_decoupled} easily allows us to understand the asymptotic behavior of $\eta_n(\lambda)$ for large $\lambda$ and $n$. Defining 
\be
\eta_{n,\infty} = \lim_{|\lambda|\rightarrow\infty}\eta_n(\lambda)\,,
\ee 
and taking the limit $|\lambda|\to\infty$ in \eqref{eq:final_decoupled}, one obtains the set of algebraic equations
\be
\eta_{n,\infty} (1+\eta_{n,\infty}) = (1+\eta_{n-1,\infty})(1+\eta_{n+1,\infty}),
\label{eq:algebraic_equation}
\ee
with $\eta_{0,\infty}=0$. It is straightforward to verify that the following ansatz satisfy \eqref{eq:algebraic_equation}
\be
\eta_{n,\infty} = \frac{(n+1)(n+2)}{2}-1\,.
\label{eq:asymptotic_condition}
\ee
As we discuss in the following, Eq.~\eqref{eq:asymptotic_condition} is recovered by our numerical solution of \eqref{eq:final_decoupled}.

\section{The post-quench steady state}\label{Sec:SteadyState}

Our strategy to numerically determine the saddle-point root densities is straightforward. First, we solve Eq.~\eqref{eq:final_decoupled} for $\eta^{(r)}_n(\lambda)$ and then we find $\rho^{(r)}_n(\lambda)$, $\rho^{(r)}_{h,n}(\lambda)$ by solving the partially decoupled Bethe equations \eqref{eq:tri-diagonal_BT2}\,--\,\eqref{eq:convention_rhot0}. For the sake of presentation, details on our numerical solution of \eqref{eq:final_decoupled} and  \eqref{eq:tri-diagonal_BT2}\,--\,\eqref{eq:convention_rhot0} are postponed to Sec.~\ref{sub:numerical}, while here we present and discuss the final result.  

\begin{figure}
\includegraphics[width=\textwidth]{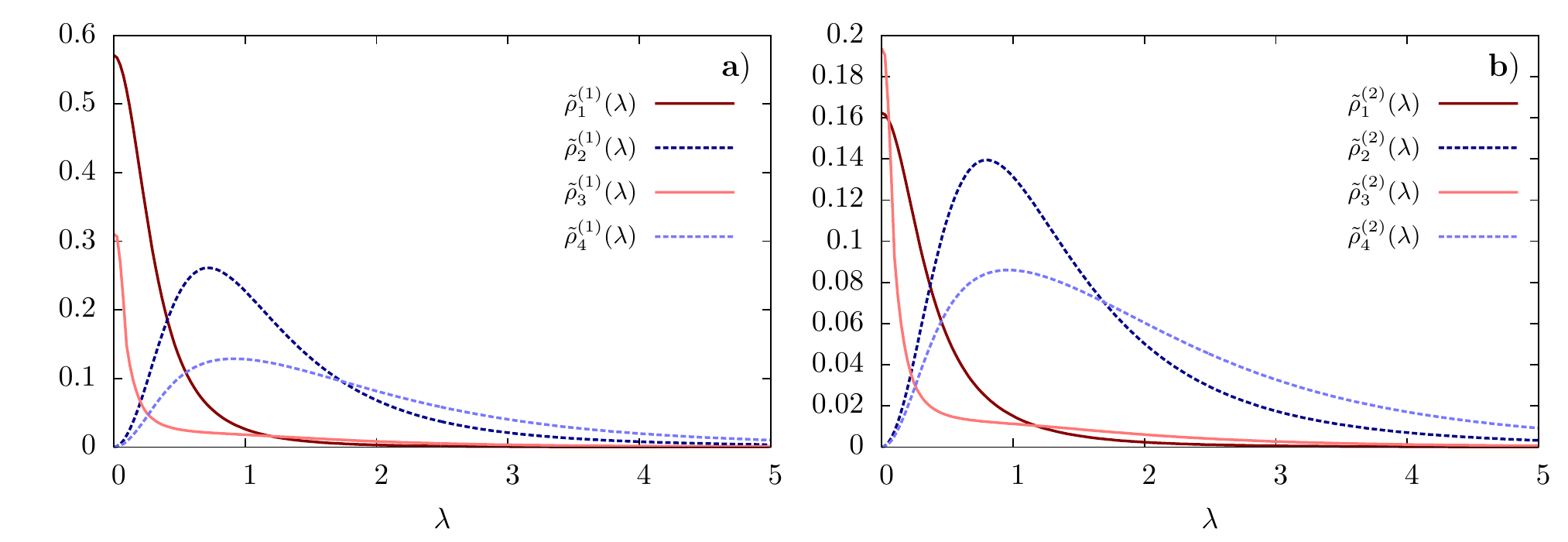}
\caption{Rapidity distribution functions of the post-quench steady state. Left and right panels show the \emph{rescaled} root densities $\tilde\rho_n^{(r)}(\lambda)$ of the first four string types $n=1,2,3,4$ for the two species of rapidities $r=1,2$ describing the eigenstates of our system. Rescaled root densities are defined as $\tilde{\rho}_n^{(r)}(\lambda)=n^2  \bar\rho_n^{(r)}$ for odd $n$ and $\tilde{\rho}_n^{(r)}=10 n^2 \bar \rho^{(r)}_n(\lambda)$ for even $n$. The rescaling is performed to show all the root densities on the same plot.}
\label{fig:densities}
\end{figure}

The rapidity distribution functions $\rho_n^{(1)}(\lambda)$, $\rho_n^{(2)}(\lambda)$ characterizing the post-quench steady state are displayed in Fig.~\ref{fig:densities}. Note that we have rescaled the rapidity distributions corresponding to bound particles, as they are significantly smaller than those of unbound particles. The knowledge of these rapidity distributions in principle allows us to compute the long-time limit of any local observable after the quench.

\begin{figure}
\includegraphics[width=\textwidth]{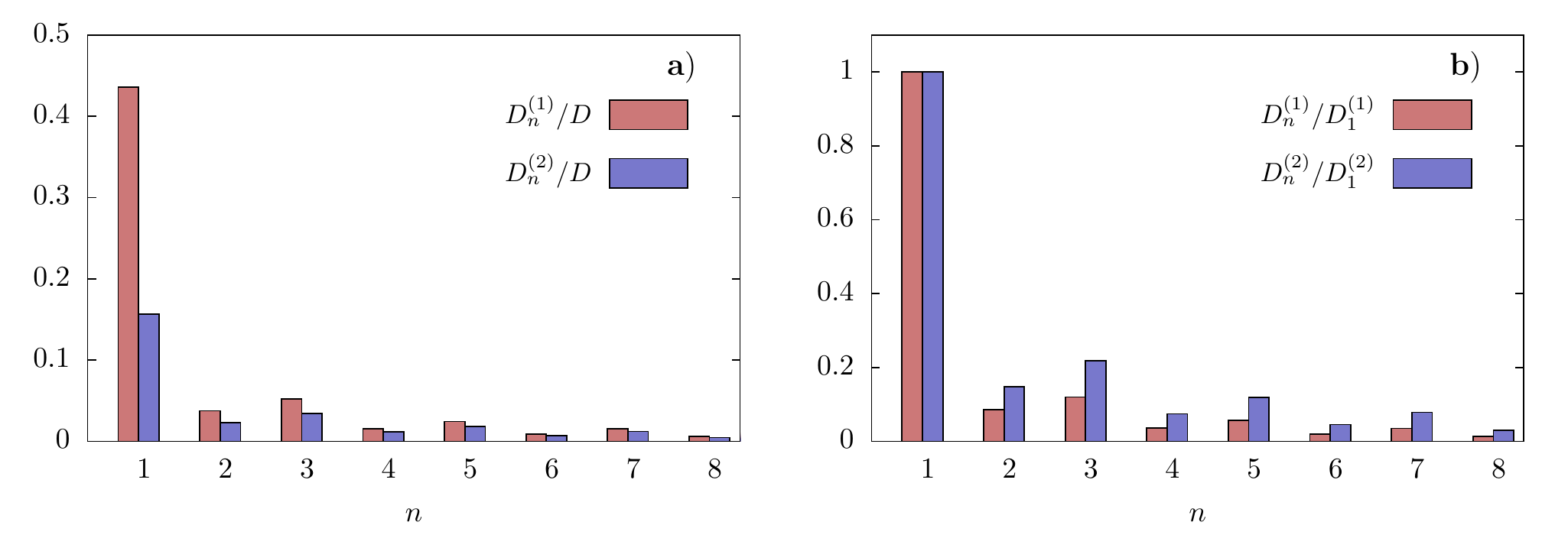}
\caption{Normalized contributions  $D_{n}^{(1)}$, $D_{n}^{(2)}$ [defined in \eqref{eq:string_content}] of the bound particles to the densities $D^{(1)}$ and $D^{(2)}$. The plots show that the density of $n$-particle bound states rapidly decreases with $n$, while the value of $D_{n}^{(2)}$ is always comparable to that of $D_{n}^{(1)}$.}
\label{fig:relative_densities}
\end{figure}

The bound-state content of the post-quench steady state is displayed in Fig.~\ref{fig:relative_densities}. The density of unbound particles is the prominent one, even though finite densities of $n$-particle bound-states are non-negligible for small $n$. Also note that the sequence of densities is not monotonic in $n$, but displays an even/odd effect.  

\begin{figure}
\includegraphics[width=\textwidth]{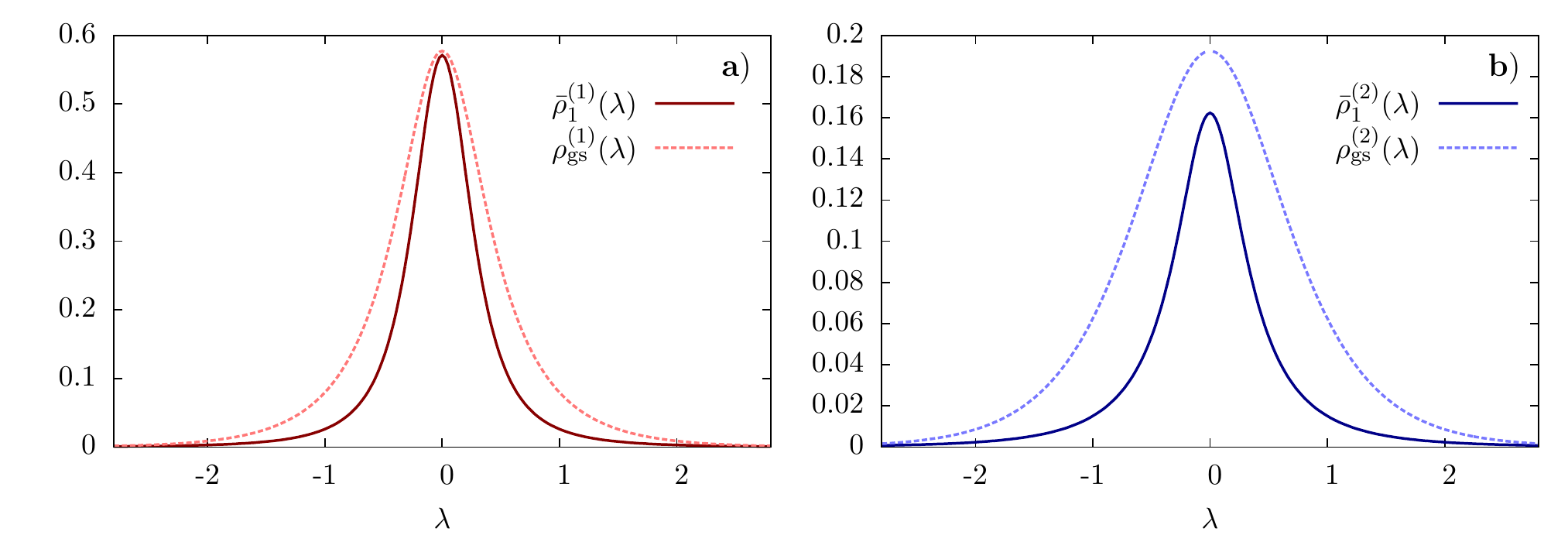}
\caption{Comparison between the rapidity distribution functions $\rho^{(1)}_1(\lambda)$, $\rho^{(2)}_1(\lambda)$ of the ground-state and the post-quench steady state. The plot shows that in the steady state the density of unbound particles is smaller, as it contains non-negligible densities of bound-states, {\it c.f.} Fig.~\ref{fig:relative_densities}}
\label{fig:ground_state_rapidities}
\end{figure}

The post-quench steady state lies in the same magnetization sector of the ground-state of the model: they both have $D^{(1)}=2/3$ and $D^{(2)}=1/3$ \cite{johannesson-86}. The ground-state, however, displays absence of bound-states so that ${\rho^{(1)}_n(\lambda)\equiv\rho^{(2)}_n(\lambda)\equiv0}$ for $n\geq 2$. A comparison between the rapidity distributions $\rho^{(1)}_1(\lambda)$, $\rho^{(2)}_1(\lambda)$ of the ground-state and the steady state is displayed in Fig.~\ref{fig:ground_state_rapidities}. We see that even though higher bound-states have small densities in the steady state they significantly influence the rapidity distribution $\rho^{(2)}_1(\lambda)$ of the second species of particles.

In the next subsection, we discuss on the computation of the local conserved operators both on the initial and the post-quench steady state. This will be crucial in order to test the validity of Eq.~\eqref{eq:final_decoupled} and the accuracy of our numerical solution. Next, we provide further details on the numerical scheme employed to solve the saddle-point integral equations.

\subsection{The local conserved charges}
\label{sub:tests}

Since our system is integrable there exists an infinite number of local and quasi-local conserved operators, or charges, commuting with the Hamiltonian. Let us focus only on the local charges and indicate them as $\{ Q_{n}\}_n$, where $Q_2=H$ by convention. Since these operators are conserved, their expectation values on the initial state and on the long-time stationary state have to be equal. This fact provides the basis for the main test of the validity of our results.

In Appendix~\ref{sec:algebraic_bethe} we derive the following expression for the expectation value of a given charge on a Bethe state
\be
\lim_{L\rightarrow\infty}\frac{1}{L}\braket{|\{\rho^{(r)}_{n}\}|  Q_m|\{\rho^{(r)}_{n}\}}=\sum_{n=1}^{\infty}\int_{-\infty}^{+\infty}{\rm d}\lambda\left( \rho^{(1)}_{n}(\lambda)c^{(1)}_{m,n}(\lambda)+\rho^{(2)}_{n}(\lambda)c^{(2)}_{m,n}(\lambda)\right)\,,
\label{eq:charge1}
\ee
where 
\bea
c^{(1)}_{m+1,n}(k)&=&(-1)^mi\frac{\partial^m}{\partial \lambda^m}\log\left[\frac{k+in/2}{k-in/2}\right]\,,\qquad\qquad m\geq 1\,,\label{eq:charge2}\\
c^{(2)}_{m+1,n}(\lambda)&\equiv& 0\,, \qquad\qquad m\geq 1\,.\label{eq:charge3}
\eea
We indicated with $\ket{\{\rho^{(r)}_{n}\}}$ a Bethe state which corresponds to the rapidity distributions $\{\rho^{(r)}_{n}\}$ in the thermodynamic limit. Note that the second species of particles does not contribute to the value of any of the local conserved charges. Equations \eqref{eq:charge1}, \eqref{eq:charge2} and \eqref{eq:charge3} immediately allow us to numerically compute, after integration, the value of the charges on the post-quench steady state. 

In order to compute the expectation value of the local charges on the initial state, we exploit the method outlined in \cite{fe-13} for the case of the XXZ spin-$1/2$ chain. As discussed in Appendix~\ref{sec:derivation_charges}, we find that the same method can straightforwardly be applied also in our case. First, we define the generating function $\Omega_{\Psi_0}(\lambda)$ such that
\be
\frac{\partial^{n}}{\partial\lambda^{n}}\Omega_{\Psi_0}(\lambda)\Big|_{\lambda=0}=\lim_{L\rightarrow\infty}\frac{1}{L}\langle \Psi_0| Q_{n+2}|\Psi_0\rangle\,.
\label{eq:def_generating_function}
\ee
Then, we show (\emph{cf}. Appendix~\ref{sec:derivation_charges}) that the generating function has the following simple expression
\be
\Omega_{\Psi_0}(\lambda)=-\frac{4  (3 + 2 \lambda^2)}{3 (3 + 7 \lambda^2 + 4 \lambda^4)}\,.
\label{eq:final_generating_function}
\ee
From \eqref{eq:final_generating_function} we immediately obtain
\be
\lim_{L\rightarrow\infty}\frac{1}{L}\langle \Psi_0|Q_{2k+1}|\Psi_0\rangle=0\,, \quad k=1,2,\ldots \infty\,,
\ee
together with the explicit expression of the even charges. As an example we report the first few charges, \emph{i.e.}, 
\bea
\lim_{L\rightarrow\infty}\frac{1}{L}\braket{\Psi_0| Q_{2}|\Psi_0}&=&-\frac{4}{3}\,, \label{eq:initial_charge2}\\
\lim_{L\rightarrow\infty}\frac{1}{L}\braket{\Psi_0| Q_{4}|\Psi_0}&=&\frac{40}{9}\,,\label{eq:initial_charge4}\\
\lim_{L\rightarrow\infty}\frac{1}{L}\braket{\Psi_0| Q_{6}|\Psi_0}&=&-\frac{736}{9}\,.
\label{eq:initial_charge6}
\eea
We can also readily write down two additional local conserved charges which are independent from the operators $Q_{n}$. These are $\mathcal{N}_{1}$ and $\mathcal{N}_{2}$ defined in \eqref{eq:n1} and \eqref{eq:n2}. The expression for their expectation value on Bethe states is given in \eqref{eq:density1} and \eqref{eq:density2}. In addition, by exploiting the simple matrix product form of the initial state, it is easy to compute that
\bea
\lim_{L\rightarrow\infty}\frac{1}{L}\braket{\Psi_0|\mathcal{N}_{1}|\Psi_0}&=&\frac{2}{3}\,, \label{eq:particle_number1}\\
\lim_{L\rightarrow\infty}\frac{1}{L}\braket{\Psi_0|\mathcal{N}_{2}|\Psi_0}&=&\frac{1}{3}\,.
\label{eq:particle_number2}
\eea
As we will discuss in the next subsection, all these values are correctly recovered by our numerical solution for the steady state rapidity distributions \eqref{eq:final_decoupled}.

\subsection{The numerical solution}
\label{sub:numerical}

The partially decoupled version of TBA integral equations is in general more convenient from the numerical point of view. Nevertheless, as a general rule we have always solved both the coupled and partially decoupled versions for all the integral equations that we have considered. The agreement between the two results gives us a useful check for our numerical methods.

All the systems considered in this work feature an infinite number of equations, therefore, to provide a numerical solution one needs to ``truncate" them retaining only a finite number $N_{\rm eq}$ of equations.  The solution to the truncated system, indicated by $X_n(\lambda)$, is then an approximation to the exact result and becomes exact in the limit $N_{\rm eq}\to\infty$. The partially decoupled version of the integral equations allows us to consider larger $N_{\rm eq}$ and usually allows to reach better accuracy. The agreement between the numerical solution of the coupled and partially decoupled equations is almost perfect for the functions $X_{n}(\lambda)$ with $n<N_{\rm eq}$ and $n$ not too close to $N_{\rm eq}$, while small discrepancies arise for $X_{n}(\lambda)$ with $n\lesssim N_{\rm eq}$ due to the effect of truncation. In the following we briefly comment on our numerical solution of the partially decoupled equations, which yielded our most accurate results.

First, we solved the saddle point equations \eqref{eq:final_decoupled} for $\eta_n^{(r)}(\lambda)$ using successive iterations. The truncation procedure followed here is analogous to that of Refs.~\cite{wdbf-14,PMWK14}. Since the driving terms $d_n(\lambda)$ in \eqref{eq:final_decoupled} are all equal for $n$ of a given parity, it is natural to expect the same even/odd effect to be present for the solutions $\eta_n(\lambda)$. In particular, one expects that for large $n$ it holds $\eta_n/\eta_{n+2}\sim 1$. More precisely, based on the asymptotic behavior \eqref{eq:asymptotic_condition}, we truncated the equations \eqref{eq:final_decoupled} using the condition
\be
\frac{\eta_{N_{\rm  Eq}+1}(\lambda)}{\eta_{N_{\rm  Eq}-1}(\lambda)}=\frac{\eta_{N_{\rm  Eq}+1,\infty}}{\eta_{N_{\rm  Eq}-1,\infty}}=1+\frac{4}{N_{\rm eq}}+O\left(\frac{1}{N_{\rm eq}^2}\right)\,.
\label{eq:boundary_condition}
\ee
We checked that for fixed $n$ and $\lambda$ the value $\eta_n(\lambda)$ converges to a well defined number increasing $N_{\rm eq}$. In addition, the solution obtained using \eqref{eq:boundary_condition} is consistent with the one of the coupled version of the integral equations \eqref{eq:final_decoupled}. 

The second step of the numerical solution is to apply the same iterative technique to the partially decoupled Bethe-Takahashi equations \eqref{eq:tri-diagonal_BT1}-\eqref{eq:tri-diagonal_BT2} determining $\rho_n^{(r)}(\lambda)$. The truncation in this case is performed in the standard fashion
\be
\rho_{t,N_{\rm  Eq}+1}(\lambda)=\rho_{t,N_{\rm  Eq}}(\lambda)\,.
\ee  
Note that, while the functions $\eta_n^{(r)}(\lambda)$ and $\rho_n^{(r)}(\lambda)$ live on the real line, their numerical realization is taken on an evenly distributed mesh of $N_\mathrm{points}$ points on a finite interval $[-\Lambda,\Lambda]$. Extrapolation of $\eta_n^{(r)}(\lambda)$ and $\rho_n^{(r)}(\lambda)$ in $N_\mathrm{points}\rightarrow \infty$ is necessary to obtain an accurate solution, while the other two parameters $\Lambda$ and $N_{\rm eq}$ are fixed to reasonably large values. The results shown in the figures are computed using $\Lambda = 50$ and $N_{\mathrm{eq}}=40$.

We tested our numerical solution by comparing the theoretical expectation values of the conserved charges \eqref{eq:initial_charge2}-\eqref{eq:initial_charge6} to the corresponding numerical results obtained via \eqref{eq:charge1}. After the extrapolation of $\rho_n^{(r)}(\lambda)$ to $N_\mathrm{points}\rightarrow \infty$, the numerical values of the charges have less than $1\%$ error. The expectation values of $\mathcal{N}_1$ and $\mathcal{N}_2$ are slightly less accurate as they receive significant contributions from higher strings whose densities are affected by the truncation in the number of  equations. Choosing the parameters as specified above, the errors in the expectation values of the particle numbers are $3.6\%$ for the first species and $6.2\%$ for the second. As an additional test of the numerical solution, we checked that the quench action \eqref{eq:SQA} evaluated at the saddle point is indeed zero, \emph{i.e.}, the two terms appearing in the difference \eqref{eq:SQA} are equal. After the extrapolation of the densities to $N_\mathrm{points}\rightarrow \infty$ we find that the difference between the overlap and the entropy term in \eqref{eq:SQA} is less than $1\%$ of each term.

\section{Entanglement Dynamics and Elementary Excitations}%
\label{Sec:entanglement}
In this section we exploit the knowledge of the post-quench stationary state, determined in the above section, to investigate 
the finite-time dynamics of the system after the quench. 
We focus on the time evolution of the entanglement entropy after the quench. 
The amount of entanglement between a subsystem $A$ and the rest of the system $\bar A$, is measured by the the entanglement entropy 
$S_A(t)$ defined as 
\be
S_A(t)=-\textrm{tr}[\rho_A(t)\log\rho_A(t)]\,,
\ee
where $\rho_A(t)$ is the time evolving density matrix of the system reduced to the subsystem $A$, \emph{i.e.} 
$\rho_A\equiv {\rm Tr}_{\bar A}|\Psi(t)\rangle\langle \Psi(t)|$. 
This entanglement entropy is known to give very important information about the system, both in and out of equilibrium, see, \emph{e.g.}, the reviews~\cite{entanglement:review, ECP:review, CCD:review, L:review}.

\begin{figure}
\includegraphics[width=0.8\textwidth]{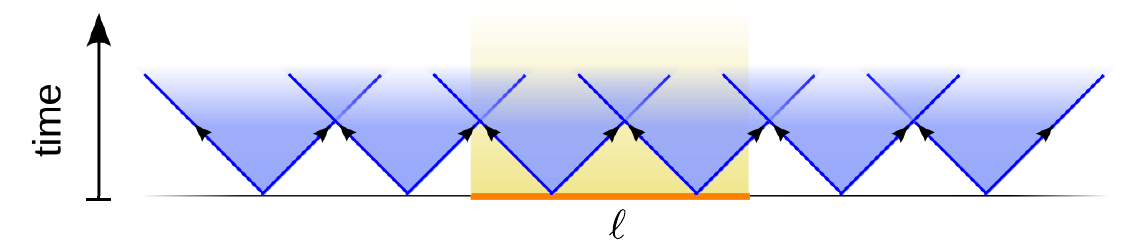}
\includegraphics[width=0.8\textwidth]{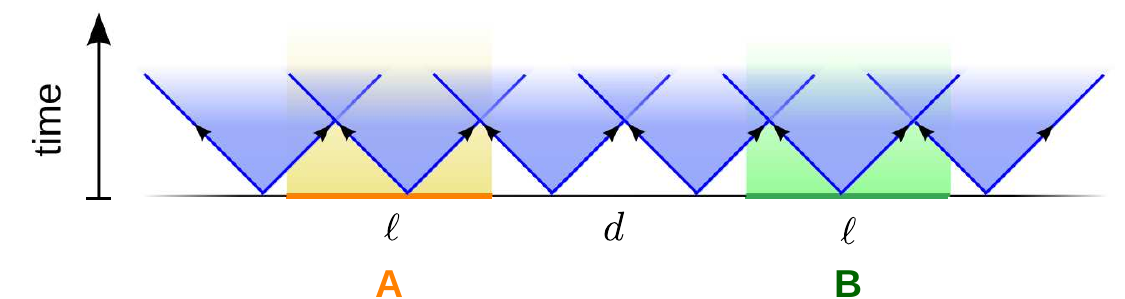}
\caption{Pictorial representation of the quasi-particle interpretation of the entanglement dynamics in two different configurations. Blue solid lines represent pairs of quasi-particles moving at the maximal velocity, while slower pairs of quasi-particles are represented by a light blue halo. The top panel depicts the evolution of entanglement entropy between a subsystem of length $\ell$ and the rest of the infinite system, detailed in Sec.~\ref{sec:entropy}. The entanglement entropy is computed by counting all the pairs of quasi-particles with one quasi-particle in the subsystem and the other in the rest.
The bottom panel shows the configuration considered in Sec.~\ref{sec:mutual}. Here we deal with two subsystems $A$ and $B$ of length $\ell$ separated by a distance $d$, and compute the mutual information $I_{A:B}(t)$ by counting all the pairs of quasi-particles with one quasi-particle in $A$ and the other in $B$. }
\label{fig:sketch}
\end{figure}

A convenient way to describe the evolution of entanglement is by means of the quasi-particle picture originally proposed in Ref.~\cite{cc-05,cc-06}. 
In essence, one postulates that the quench creates pairs of correlated quasi-particles in any spacial point of the system. 
Only pairs created at the same point are correlated and carry entanglement through the system. 
For $t>0$, the quasi-particles move ballistically in opposite directions  and, as a consequence of momentum conservation, 
the two correlated quasi-particles have opposite velocities $\pm v(\lambda)$, where $\lambda$ is the rapidity parametrising the dispersion relation. 
When moving through the system, the quasi-particles correlate regions which were initially uncorrelated as pictorially shown in Fig.~\ref{fig:sketch}. 
The entanglement entropy $S_A$ is then a weighted (by a function $s(\lambda)$) measure of the number of pairs with one quasi-particle in 
$A$ and the other in $\bar A$. 
It has been shown in many non-interacting models~\cite{NR:entharmonicoscillators, FC:ent, CTC:entanglement, EP:entperiodicquench,chmm-16}, that the predictions of the this quasi-particle picture become \emph{exact} in the space-time scaling limit of large times and subsystem sizes.
Furthermore the qualitative picture for the entanglement entropy evolution has been shown to be correct even in 
numerical simulations of many interacting integrable and non-integrable models, as \emph{e.g.} in Refs.  \cite{dmcf-06,lk-08,kh-13,cce-15,nahum-17}.
The same picture also provides the entanglement dynamics in local quenches \cite{cc-07l,ep-07,sd-11,B:localquench} 
and in inhomogeneous situations \cite{dsvc17,a-17}.

In Ref.~\cite{ac-16} it has been shown that the quasi-particle picture gives, in the space-time scaling limit, exact predictions 
even for interacting integrable models, provided that an appropriate choice for the functions $v(\lambda)$ and $s(\lambda)$ is made. 
One has to introduce multiple species of quasi-particles moving at the velocities $v_n(\lambda)$, which are the group velocities of 
elementary excitations over the stationary state described by  $\{\bar\rho_n(\lambda)\}$. 
In interacting models the velocities $v_ n(\lambda)$ are generically state-dependent and non-trivially encode the effects of the 
interactions -- they fulfil a set of integral equations depending on $\{\bar\rho_n(\lambda)\}$. 
In the non-interacting limit they reduce to the bare velocities $v^0_n(\lambda)=\varepsilon_ n'(\lambda)/p_ n'(\lambda)$, 
where $\varepsilon_n(\lambda)$ and  $p_n(\lambda)$ are respectively the bare energy and momentum. 
The choice of the quasi-particles' velocities is in agreement with the one found in transport 
problems~\cite{BCDF:transport, CADY:hydro, BF:defect} and it is ultimately related to the fundamental 
observation~\cite{BEL:lightcone} that the group velocities of the elementary excitations are the relevant velocities for the propagation 
of information in interacting integrable models. 
The natural choice for the weighting functions $s_n(\lambda)$ is to set them equal to the Yang-Yang entropy density per rapidity and species. 
This choice guarantees that the extensive parts of entanglement entropy and thermodynamic entropy coincide at infinite times, 
in agreement with some
general expectations \cite{santos-2011,gurarie-2013} as well as  analytical findings in non-interacting models~\cite{FC:ent,CKC:TGLL,KBC:Ising}.

In our case, using the quasi-particle interpretation, we can write the entanglement entropy between a subsystem of contiguous spins of 
length $\ell$ and the rest of the system as 
\be
S_\ell(t)=\sum_{r=1,2}\,\sum_{n=1}^{\infty}\,\int\!\!{\rm d}\lambda\,\, s_{n}^{(r)}(\lambda)\left\{2 t |v_{n}^{(r)}(\lambda)|\,\theta_{\rm H}(\ell-{2|v_{n}^{(r)}(\lambda)|t})+\ell\, \theta_{\rm H}({2|v_{n}^{(r)}(\lambda)|t}-\ell)\right\}\,.
\label{eq:entanglemententropy}
\ee
Here we used that the indices labeling quasi-particles are $n=1,2,\ldots$ and $r=1,2$. The Yang-Yang entropy density $s_{n}^{(r)}(\lambda)$ appearing in \eqref{eq:entanglemententropy} is given by  
\be
 s_{n}^{(r)}(\lambda)=\left(\rho_n^{(r)}(x)+\rho_{h,n}^{(r)}\right)\ln\left(\rho_n^{(r)}(x)+\rho_{h,n}^{(r)}\right)-\rho_{n}^{(r)}\ln\rho_{n}^{(r)}-\rho_{h,n}^{(r)}\ln\rho_{h,n}^{(r)}\,,
\ee
and velocities $v_n^{(1)}(\lambda)$ and $v_n^{(2)}(\lambda)$ fulfil the integral equations
\bea
\rho^{(2)}_{t,n}(\lambda) v^{(2)}_n(\lambda)&=&\sum_{k}\left(b_{n,k}\ast v^{(1)}_k \rho^{(1)}_k\right)(\lambda)-\sum_{k}\left(a_{n,k}\ast v^{(2)}_k \rho^{(2)}_k\right)(\lambda)\label{eq:velocities1}\\
\rho^{(1)}_{t,n}(\lambda)v^{(1)}_n(\lambda)&=&\frac{1}{2\pi}\varepsilon^{\prime}_n(\lambda)-\sum_k \left( a_{n,k}\ast v^{(1)}_k\rho_k^{(1)}\right)(\lambda)+\sum_k \left(b_{n,k} \ast v^{(2)}_k \rho_k^{(2)}\right)(\lambda)\,.
\label{eq:velocities2}
\eea
Before discussing the predictions of \eqref{eq:entanglemententropy} for the entaglement dynamics we briefly sketch the derivation of Equations \eqref{eq:velocities1}\,--\,\eqref{eq:velocities2}, and refer the reader to Appendix~\ref{app:velocities} for the detailed calculation. 

\subsection{Velocities of the Excitations}\label{sec:velocities}

Elementary excitations over the stationary state  $\{\bar \rho_n^{(r)}(\lambda)\}$ are constructed by considering a microscopic representation of the stationary state, described by a configuration of integers $\{I_\alpha^n,J_\beta^m\}$ (\emph{cf}.~\eqref{eq:logarithmicBTE}), and adding or removing one integer. The operation of adding such an integer to those in the sector labelled by $n$ and $r$, which we call \emph{excitation of string type $n$ and species $r$}, can be interpreted as the addition of a quasi-particle with rapidity $\lambda$, energy $\varepsilon_n^{d\,(r)}(\lambda)$ and momentum $p_n^{d\,(r)}(\lambda)$. The latter quantities are given by  
\bea
\varepsilon^{d\,(a)}_n(x) &=& \varepsilon_n(x)\delta_{a,1}+\sum_k\!\!\int\!{\rm d}z\,\,\varepsilon^{\prime}_{k}(z)\vartheta^{(1)}_k(z)F^{1a}_{kn}(z|x)\,,\qquad\qquad\qquad a=1,2\,,\label{eq:dresseeps}\\
p^{d\,(a)}_n(x) &=& p_n(x)\delta_{a,1}+\sum_k\!\!\int\!{\rm d}z\,\,p^{\prime}_{k}(z)\vartheta^{(1)}_k(z)F^{1a}_{kn}(z|x)\,,\qquad\qquad\qquad a=1,2\,.\label{eq:dressedp}
\eea
The shift functions $F^{11}_{kn}(z|x)$ and $F^{12}_{kn}(z|x)$ appearing here are found by solving a system of integral equations, reported in Equations \eqref{eq:F11}\,--\,\eqref{eq:F22} of Appendix~\ref{app:velocities}. The quantities $\varepsilon_n^{d\,(r)}(\lambda)$ and $p_n^{d\,(r)}(\lambda)$ are called dressed energy and momentum. They differ from the bare ones $\varepsilon_n^{(r)}(\lambda)$ and $p_n^{(r)}(\lambda)$ because of the effects of the interaction encoded in  $\{F^{1a}_{kn}(z|x)\}_{a=1,2}$. All this has a very simple physical interpretation -- adding a new quasi-particle to the system has a feedback on all the others because of the interaction, so its effective dispersion relation changes. 

Given the dispersion relation of an excitation, \emph{i.e.} its dressed energy and momentum, we can find its group velocity from the formula
\be
v_n^{(r)}(\lambda)\equiv\frac{{\rm d}\varepsilon_n^{d\,(r)}(\lambda)}{{\rm d}p_n^{d\,(r)}(\lambda)}=\frac{\varepsilon^{d\,(r)\,\prime}_n(\lambda)}{p^{d\,(r)\,\prime}_n(\lambda)}\,.
\label{eq:defvelocity}
\ee
Taking the derivative of \eqref{eq:dresseeps} and \eqref{eq:dressedp} and using \eqref{eq:F11}\,--\,\eqref{eq:F22} we find that
\be
p^{d\,(r)\,\prime}_n(\lambda)=2\pi \rho_{t,n}^{(r)}(\lambda)\,,
\ee 
and that $\{\varepsilon^{d\,(r)\,\prime}_n(\lambda)\}_{r=1,2}$ fulfil the following system of integral equations
\bea
\varepsilon^{d\,(2)\,\prime}_n(\lambda)&=&\sum_{k}\left(b_{n,k}\ast \varepsilon^{d\,(1)\,\prime}_k \right)(\lambda)-\sum_{k}\left(a_{n,k}\ast \varepsilon^{d\,(2)\,\prime}_k\right)(\lambda)\,,\\
\varepsilon^{d\,(1)\,\prime}_n(\lambda)&=&\varepsilon^{\prime}_n(\lambda)-\sum_k \left( a_{n,k}\ast \varepsilon^{d\,(1)\,\prime}_k \right)(\lambda)+\sum_k \left(b_{n,k} \ast \varepsilon^{d\,(2)\,\prime}_k\right)(\lambda)\,.
\eea
Substituting the definition \eqref{eq:defvelocity} readily gives the system \eqref{eq:velocities1}\,--\,\eqref{eq:velocities2}.

\begin{figure}
\includegraphics[width=\textwidth]{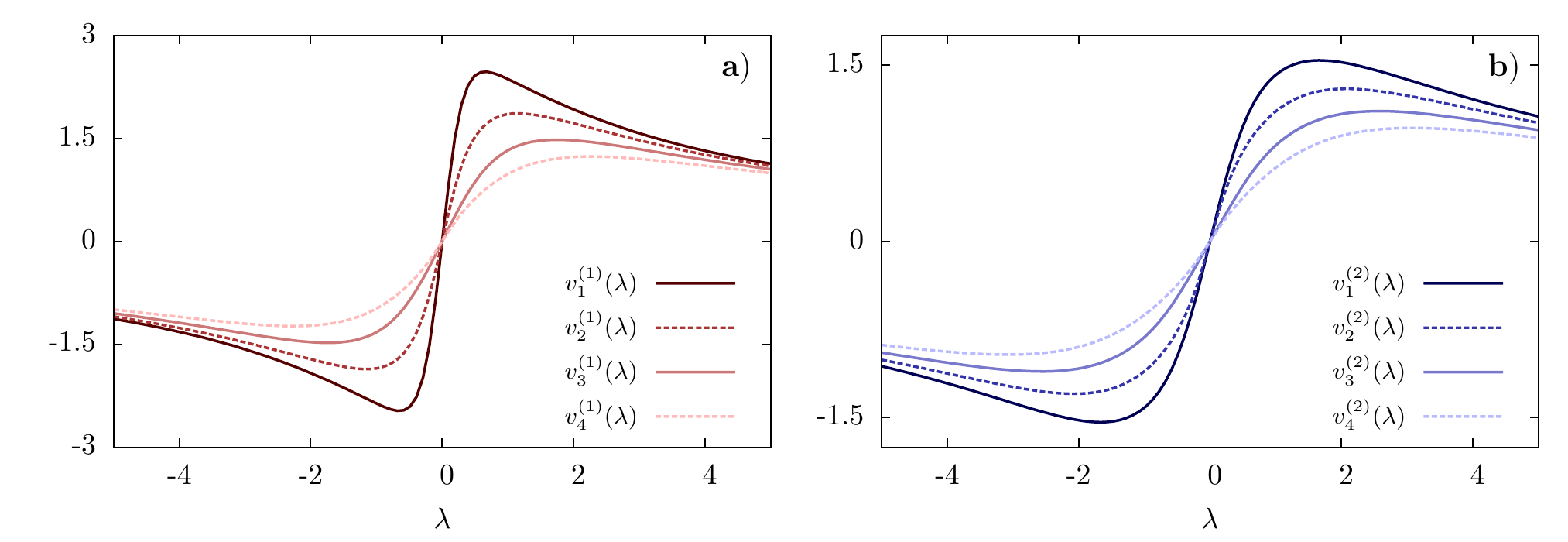}
\caption{Velocities of the elementary excitations over the stationary state $\{\bar\rho_{n}^{(r)}(\lambda)\}$ (\emph{cf}. Eq.~\eqref{eq:general_saddle_point}). Left and right panels show the velocities of elementary excitations of string type $n=1,2,3,4$, respectively of the first and second species ($r=1,2$).}
\label{fig:velocities2}
\end{figure}

For the the purpose of the numerical solution, it is convenient to follow the steps illustrated in Appendix~\ref{sec:general_calculations} and write the system \eqref{eq:velocities1}\,--\,\eqref{eq:velocities2} in a partially decoupled form 
\bea
\rho_{t,n}^{(1)}(\lambda)v^{(1)}_n(\lambda) &=& -s^{\prime}(\lambda)\delta_{n,1}+s\ast\left( \rho_{h,n-1}^{(1)}v^{(1)}_{n-1}+\rho_{h,n+1}^{(1)}v^{(1)}_{n+1}\right)(\lambda)+s\ast \rho_{n}^{(2)}v^{(2)}_{n}(\lambda)\,,\label{eq:vel1}\\
\rho_{t,n}^{(2)}(\lambda) v^{(2)}_{n}(\lambda)&=& s\ast\left( \rho_{h,n-1}^{(2)}v^{(2)}_{n-1}+\rho_{h,n+1}^{(2)}v^{(2)}_{n+1}\right)(\lambda)+s\ast \rho_{n}^{(1)}v^{(1)}_{n}(\lambda)\,.\label{eq:vel2}
\eea
These integral equations can be readily solved numerically. In Fig.~\ref{fig:velocities2} we report the velocities of the two species of elementary excitations with the first four string types, constructed over the stationary state $\{\bar\rho_{n}^{(r)}(\lambda)\}$ (\emph{cf}. Eq.~\eqref{eq:general_saddle_point}). From the plot we see that the velocities are odd functions of $\lambda$ with a minimum and a maximum reached for finite values of $\lambda$. The maximal velocity for the propagation of information is given by that of excitations of the first species and string type $1$, $v_{\rm max}=\max_\lambda v_1^{(1)}(\lambda)$. The maximal velocity for excitations of the second species is given by $v_{\rm max}^{(2)}=\max_\lambda v_1^{(2)}(\lambda)$ with $v_{\rm max}^{(2)}\approx 0.5\, v_{\rm max}$.

\subsection{Entanglement Dynamics} 
\label{sec:entropy}
\begin{figure}[t]
\includegraphics[width=\textwidth]{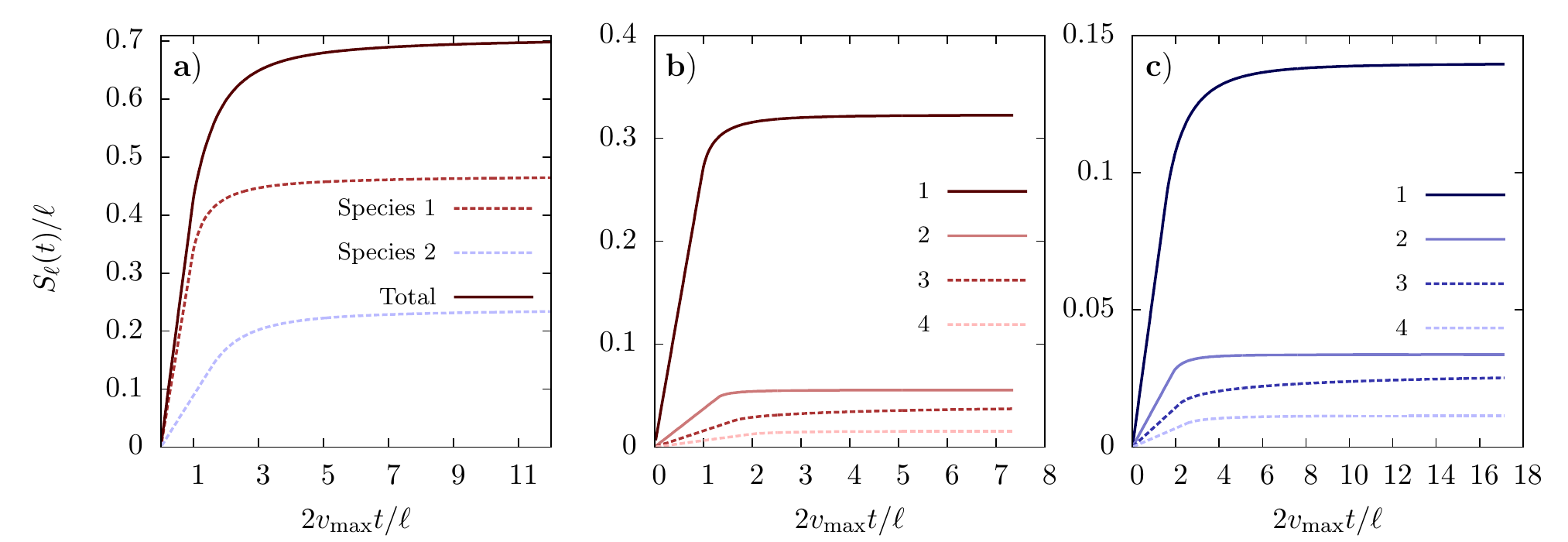}
\caption{Entanglement evolution after a quench from the initial state \eqref{eq:initial_state}. The left panel shows the time evolution of the entanglement entropy as a function of $2 v_{\rm max} t/ \ell$ (magenta solid line), it also shows the separate contributions carried by quasi-particles of the first (red dashed line) and second species (blue dashed line). The central and right panels show the string-type  resolved ($n=1,2,3,4$) contributions to the entanglement dynamics, respectively for the first and the second species of excitations.}
\label{fig:entanglement}
\end{figure}

Let us now consider the entanglement dynamics predicted by Eq.~\eqref{eq:entanglemententropy}. 
In Fig.~\ref{fig:entanglement}a, we report the time evolution of the entanglement entropy after a quench from the initial state \eqref{eq:initial_state}. 
As is customary, we plot $S_\ell(t)/\ell$ as a function of the scaling variable $2 v_{\rm max}t/\ell$. 
The plot clearly shows the standard spreading of entanglement entropy \cite{cc-05}: 
there is a linear increase of the entanglement entropy for $t< \ell/2v_{\rm max}$ governed by the fastest quasi-particles, 
followed by a slow saturation dictated by all the other slower quasi-particles. 
For the initial state considered, the largest contribution to the entanglement is coming from the fastest quasi-particles: those of species $r=1$ and string type $n=1$
(\emph{cf.} Fig. \ref{fig:relative_densities}).
This observation is confirmed by the species resolved lines in Fig.~\ref{fig:entanglement}a,  which show that the particles of the first species bring almost twice as much entanglement as those of the second species. A further confirmation comes from the string-type resolved plots in Figs~\ref{fig:entanglement}b and \ref{fig:entanglement}c, which demonstrate that the contribution of bound states is strongly suppressed.
A final observation is that the asymptotic value of the entanglement entropy is approximately 0.7 which is smaller than $\log 3\approx 1.1$, 
the maximal density of entropy per site in the spin-$1$ chain (indeed $\log 3$ corresponds to the density of thermodynamic entropy in 
the infinite temperature state). 

\subsection{Mutual Information}
\label{sec:mutual}

The entanglement entropy is not the ideal quantity to highlight the contribution of all the different quasi-particles: 
the contribution of the quasi-particles bringing more entanglement covers all the others. 
The contribution of different quasi-particles can be resolved more effectively considering two disjoint intervals. 
Let us take two spin blocks $A$ and  $B$ of length $\ell$, separated by a distance $d$, as depicted in Fig.~\ref{fig:sketch}, and 
focus on the \emph{mutual information} 
\be
I_{A:B}=S_A+S_B-S_{A\cup B}
\ee
between the two subsystems $A$ and $B$. 
The mutual information after a quench is also believed to signal the non-integrability and chaotic behavior of a system 
via the breakdown of the quasi-particle picture \cite{abgh-15,sm-15}. 

In the quasi-particle picture, the mutual information counts all the pairs of quasi-particles with one quasi-particle in $A$ and the other in 
$B$, as pictorially shown in Fig.~\ref{fig:sketch}. Its time evolution is then simply written down generalizing the result of \cite{ac-16} to 
two species of excitations   
\begin{align}
I_{A:B}(t)&=\sum_{r=1,2}\sum_{n=1}^\infty \int\!\!\mathrm{d}\lambda\,\biggl[\left(2|v^{(r)}_n(\lambda)| t - d\right)\chi_{[d,d +\ell]}(2 |v^{(r)}_n(\lambda)| t)\notag\\
&\qquad\qquad\qquad\qquad+ \left(d + 2 \ell - 2 |v^{(r)}_n(\lambda)| t\right)\chi_{[d+\ell,d +2\ell]}(2 |v_n^{(r)}(\lambda)| t)\biggr]s_n^{(r)}(\lambda)\,,
\end{align}
where $\chi_{[a,b]}(x)$ is the characteristic function of $[a,b]$, \emph{i.e.} it is equal to $1$ if $x\in[a,b]$ and equal to $0$ otherwise. 

\begin{figure}
\includegraphics[width=\textwidth]{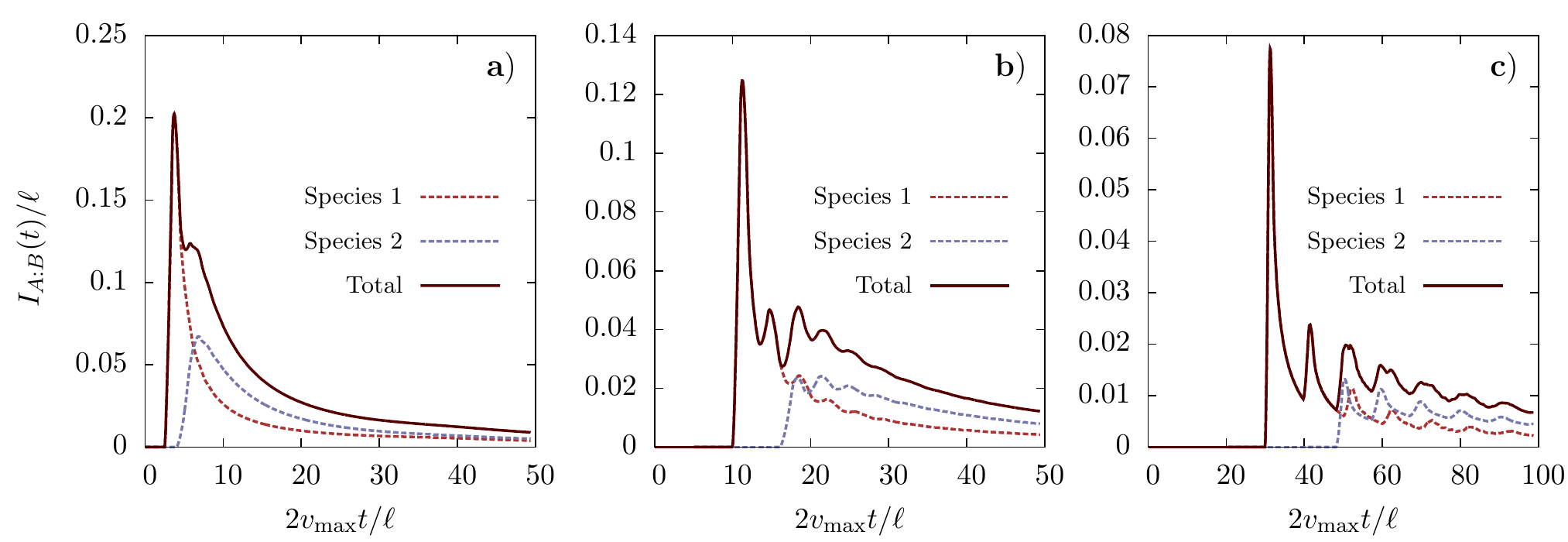}
\caption{Time evolution of the mutual information starting from the initial state \eqref{eq:initial_state}. The three panels correspond to increasing values of the ratio $d/\ell$, respectively $2.5,10,30$. The plots show the total mutual information (magenta solid line) together with the separate contributions carried by quasi-particles of the first (red dashed line) and second species (blue dashed line).}
\label{fig:mutual}
\end{figure}

The time evolution of the mutual information is reported in Fig.~\ref{fig:mutual}, where we plot $I_{A:B}(t)/\ell$ as a function of $2 v_{\rm max}t/\ell$ for three different values of the ratio $d/\ell$. We see that the contributions of different quasi-particles are easily detected as they give rise to peaks in $I_{A:B}(t)$ -- the peak due to the quasi-particles of species $r$ and string type $n$ is appearing at approximately  
\be
t_{n}^{(r)}= \frac{d+\ell}{2 v_{\max, n}^{(r)}}=  \frac{v_{\max}}{v_{\max, n}^{(r)}} t^{(1)}_{1}\,,
\ee
where we introduced $v_{\max, n}^{(r)}=\max_\lambda v_{n}^{(r)}(\lambda)$. Once again, the most prominent peak corresponds to the fastest quasi-particles, as they carry the largest amount of correlations. As shown in Figs.~\ref{fig:mutual}a\,--\, \ref{fig:mutual}c, increasing the ratio $d/\ell$ we can increase $2 v_{\rm max} t^{(1)}_1/\ell$ and separate the peaks, in this way it is easier to discern the contribution of different quasi-particles.

\section{Conclusions}\label{Sec:Conclusions}

In this paper we studied a quantum quench in a nested Bethe-ansatz solvable model, the spin-$1$ Lai-Sutherland chain. The thermodynamic description of the latter is in terms of two different species of quasi-particles, each one forming an infinite number of bound states. We considered a simple initial matrix product state for which the overlaps with Bethe states are known~\cite{dkm-16} and determined the corresponding post-quench steady state by means of the Quench Action approach. We tested the validity of our result by checking that the expectation value of all the local conserved charges on the initial and long-time steady state are equal. Finally, we investigated the post-quench entanglement dynamics using a recently proposed conjecture for interacting integrable models ~\cite{ac-16}. Importantly, due to the simple structure of our initial state, all our predictions for the entanglement dynamics can in principle be tested against efficient numerical methods such as tDMRG or iTEBD, on the same lines of what done in Ref.~\cite{ac-16}.    

Our work raises a number of compelling questions. From the physical point of view, it is natural to wonder how the two different species of excitations affect the dynamics of local correlations after the quench. It is clear from our results that each species carries a non-vanishing contribution on entanglement dynamics. As a consequence, it is not possible to individually probe the dynamics of the single species by computation of entanglement entropy or mutual information. An interesting question then is whether some appropriate local correlators can be found, such that their dynamics is entirely determined by one and only one of the two species. This would result in a mechanism of effective separation in the spreading of correlations after the quench.

From the technical point of view, an interesting question stems from the structure of the overlap formula of our initial state. Since the seminal works \cite{kp-12,pozsgay-14,dwbc-14}, determinant expressions of the form \eqref{eq:overlap_formula} have appeared in several cases in the past few years for a variety of models. It is even more surprising to encounter a formula of this kind in the nested spin chain studied in this work, possibly signalling the presence of an hidden structure common to all Bethe ansatz integrable models. Characterising the states for which such a formula exists is an important open problem, with far reaching consequences, both from the purely theoretical and practical point of view.  

Our work also calls for a systematic analysis of the quasi-local charges in the spin-$1$ Lai-Sutherland chain. 
This is a necessary step towards the explicit construction of the complete GGE in this model. Such a construction would in turn lead to the possibility of considering a wider set of initial states, overcoming the technical difficulty of the computation of the overlaps required by the Quench Action approach. The explicit calculations presented in this work for the initial state \eqref{eq:initial_state}, provide a strict test for any future construction of the complete GGE.

Finally, it would be highly desirable to find explicit formulae for calculating expectation values of local operators on the stationary state in the spin-$1$ Lai Sutherland chain. Such formulae are currently known only for few non-nested Bethe ansatz integrable models, namely the Lieb-Liniger model~\cite{KCI:correlatorsLL11, P:correlatorsLL11}, the XXZ spin-$1/2$ chain~\cite{MP:correlatorsXXZ14, P:correlatorsXXZ16}, and the sinh-Gordon field theory~\cite{BePC16, NS:sinhgordon}. A promising route is to first consider expectation values at finite temperature 
and then generalize them to arbitrary excited eigenstates. This approach has been found fruitful in the case of the XXZ spin-$1/2$ chain and of the sinh-Gordon field theory.

\section{Acknowledgements}

We are grateful to Fabian Essler for inspiring discussions. BB and PC acknowledge the financial support by the ERC under Starting Grant 279391 EDEQS.

\appendix

\section{The algebraic Bethe ansatz}\label{sec:algebraic_bethe}

In this section we briefly sketch the algebraic Bethe ansatz analysis of the Hamiltonian \eqref{eq:hamiltonian}, which also allows for a systematic derivation of higher local conservation laws.

The fundamental object of the algebraic Bethe ansatz is the so called $R$-matrix, which in our case reads
\be	
R_{12}(\lambda)=\frac{\lambda}{\lambda+i} + \frac{i}{\lambda+i} \mathcal{P}_{12},
\label{eq:r_matrx}
\ee
where $\mathcal{P}_{12}$ is the permutation matrix exchanging the local spaces $h_{1}$, $h_{2}$.  The $R$-matrix \eqref{eq:r_matrx} is invariant under $SU(3)$ and it is a simple exercise to show that it satisfies
\bea
R_{12}(0)&=&\mathcal{P}_{12}\,\\
R_{12}(u)R_{21}(-u)&=&{\rm id}\,,
\eea
where ${\rm id}$ is the identity operator on $h_{1}\otimes h_{2}$. Next, introducing the Lax operator
\be
L_{0j}(\lambda)=R_{0j}(\lambda)\,,
\ee
one can define the family of transfer matrices
\be
{\tau}(\lambda)={\rm tr}\left\{L_{0N}(\lambda)\ldots L_{01}(\lambda)\right\}\,,
\label{eq:transfer_matrix}
\ee
which depend on the spectral parameter $\lambda$. Crucially, different transfer matrices commute
\be
\left[{\tau}(\mu),{\tau}(\lambda)\right]=0\,,
\ee
and as a consequence one can define a sequence of commuting operators as
\be
{Q}_{n+1}=i\frac{\partial^{n}}{\partial\lambda^n}\ln\tau(\lambda)\Bigr|_{\lambda=0}\,,\qquad\qquad\qquad\qquad n=1,\ldots, \infty\,.
\label{eq:charges_definition}
\ee 
These are usually called charges and can be written down explicitly by exploiting the properties of the $R$-matrix, as nicely explained in \cite{faddeev-96}. However, the actual expression becomes increasingly unwieldy for many practical purposes as $n$ increases.

The importance of this algebraic construction lies in the fact that the first charge is equal to the Hamiltonian \eqref{eq:hamiltonian}, namely
\be
{Q}_2=H\,.
\ee
Hence, the transfer matrix \eqref{eq:transfer_matrix} can be thought of as the generator of the local conservation laws of the model. 

As we have mentioned in Section~\ref{sec:solution}, the knowledge of the rapidities corresponding to a given eigenstate also allows for a direct calculation of expectation values of local charges. Indeed, the analytic expression of the eigenvalue of the transfer matrix \eqref{eq:transfer_matrix} on a Bethe state with rapidities $\{k_j\}$, $\{\lambda_j\}$ is known \cite{kr-81}, and one can explicitly write
\be
\tau(\lambda)\ket{\{k_j\},\{\lambda_j\}}=\nu(\{k_j\},\{\lambda_j\},\lambda)\ket{\{k_j\},\{\lambda_j\}}\,,
\label{eq:transfer_matrix_eigenvalue}
\ee
where 
\bea
\nu\left(\{k_j\},\{\lambda_j\},\lambda\right)&=&\left[a(\lambda)\right]^L\prod_{j=1}^{N}\frac{1}{a(\lambda-k_j+i/2)}\nu_1\left(\{k_j\},\{\lambda_j\},\lambda\right)+\prod_{j=1}^N\frac{1}{a(k_j-i/2-\lambda)}\,,\label{eq:transfer_eigenvalue}
\eea
and
\bea
\nu_1(\{k_j\},\{\lambda_j\},\lambda)&=&\prod_{j=1}^Na(\lambda-k_j+i/2) \prod_{r=1}^{M}\frac{1}{a(\lambda-\lambda_r+i/2)}+\prod_{r=1}^{M}\frac{1}{a(\lambda_r-i/2-\lambda)}\,,
\eea
with
\bea
a(\lambda)&=&\frac{\lambda}{\lambda+i}\,.
\eea
Then, it is straightforward to compute the action of a charge $ Q_n$ on the state $\ket{\{k_j\},\{\lambda_j\}}$: from the definition \eqref{eq:charges_definition}, the calculation is reduced to taking derivatives of the expression \eqref{eq:transfer_eigenvalue}. This calculation is completely analogous to the corresponding one in the spin-$1/2$ XXX chain and reads as 
\be
\label{eq:finitesizeQ}
\braket{\{k_j\},\{\lambda_j\}| Q_m|\{k_j\},\{\lambda_j\}}=\sum_{j=1}^{N} (-1)^mi\frac{\partial^m}{\partial \lambda^m}\log\left[\frac{\lambda+i/2}{\lambda-i/2}\right]\biggl|_{\lambda=k_j}\,.
\ee
In particular, note that one can neglect the first term in \eqref{eq:transfer_eigenvalue} since
\be
\frac{\partial^{m}}{\partial\mu^m}\left[a(\mu)\right]^{L}\Big|_{\mu=0}=0\,,\qquad\qquad m< L\,.
\ee
It is also possible to take the thermodynamic limit and evaluate the expectation value of conserved charges on Bethe states corresponding to the distributions $\rho_{n}^{r}(\lambda)$, $r=1, 2$. Starting from the expression at finite size, making use of the string hypothesis and taking the thermodynamic limit, one finally obtains
\be
\lim_{L\rightarrow\infty}\frac{1}{L}\braket{\{\rho^{(r)}_{n}\}| Q_m|\{\rho^{(r)}_{n}\}}=\sum_{n=1}^{\infty}\int_{-\infty}^{+\infty}{\rm d}\lambda\left( \rho^{(1)}_{n}(\lambda)c^{(1)}_{m,n}(\lambda)+\rho^{(2)}_{n}(\lambda)c^{(2)}_{m,n}(\lambda)\right)\,,
\ee
where 
\bea
c^{(1)}_{m+1,n}(\lambda)&=&(-1)^mi\frac{\partial^m}{\partial \lambda^m}\log\left[\frac{\lambda+in/2}{\lambda-in/2}\right]\,,\qquad\qquad m\geq 1\label{eq:final_c1}\\
c^{(2)}_{m+1,n}(\lambda)&\equiv& 0\,, \qquad\qquad m\geq 1\,,\label{eq:final_c2}
\eea
while we indicated with $\ket{\{\rho^{(r)}_{n}\}}$ a Bethe state which corresponds to the rapidity distributions $\{\rho^{(r)}_{n}\}$ in the thermodynamic limit. Remarkably, we see from \eqref{eq:finitesizeQ} and \eqref{eq:final_c2} that the second species of quasi-particles does not contribute to any of the higher local conserved charges.

\section{Bethe equations in the thermodynamic limit}\label{sec:general_calculations}

In this appendix we derive the partially decoupled form of the Bethe-Takahashi equations \eqref{eq:tri-diagonal_BT1}, \eqref{eq:tri-diagonal_BT2} following the standard Bethe ansatz literature \cite{takahashi-99}. We also introduce a series of identities which are used throughout this work.

We start by explicitly writing down the Fourier transforms 
\bea
\mathcal F\left[{a}_n(x)\right](k)&=& \int_{-\infty}^{\infty}\!\frac{{\rm d}x}{2\pi}\,\,\frac{n e^{i k x}}{x^2+n^2/4}=e^{-\frac{n|k|}{2}}\,,\\
\mathcal F\left[s(x)\right](k)&=& \int_{-\infty}^{\infty}\!{{\rm d}x}\,\,\frac{e^{i k x}}{2\cosh(\pi x)}=\frac{1}{2\cosh\left(k/2\right)}\,,
\label{eq:fourier_s}
\eea
where $s(\lambda)$ and $a_n(\lambda)$ are defined in \eqref{eq:s_function} and \eqref{eq:a_function} respectively. These allow us to show, by the convolution theorem, the identity
\be
s\ast \left(a_{j-1}+a_{j+1}\right)(\lambda)=a_j(\lambda)\,,
\label{eq:s-a_identity}
\ee
where $\ast$ denotes the convolution as in \eqref{eq:convolution}. Eq.~\eqref{eq:s-a_identity} is valid for $j\geq 1$ provided that one defines $a_{0}(x)=\delta(x)$, with $\delta(x)$ being the Dirac delta function. Defining now
\bea
A_{n,m}(\lambda)&=&\delta_{nm}\delta(\lambda)+(1-\delta_{nm})a_{|n-m|}(\lambda)+2a_{|n-m|+2}(\lambda)+\ldots+2a_{n+m-2}(\lambda)+a_{n+m}(\lambda)\,,
\label{eq:A_nm}
\eea
one can also easily show
\be
s\ast A_{nm}(\lambda)= b_{nm}(\lambda)\,,
\label{eq:identity_2}
\ee
where $b_{nm}$ is defined in \eqref{eq:b_mn}. The above identities allow us to rewrite the Bethe-Takahahshi equations \eqref{eq:TBAexplicit1}\,--\,\eqref{eq:TBAexplicit2} in compact notation as 
\bea
\rho_{h,n}^{(r)}(\lambda)+\sum_{m=1}^{+\infty}\sum_{s=1}^{2}A_{n,m}\ast C_{r,s}\ast \rho_{m}^{(s)}(\lambda)=a_{n}(\lambda)\delta_{r,1}\,.
\label{eq:thermo_bethe_equations}
\eea
Here $C_{r,s,}$ is defined by the components of the $2\times 2$ matrix
\be
C(\lambda)=
\left(
\begin{array}{cc}
\delta(\lambda)&-s(\lambda)\\
-s(\lambda)&\delta(\lambda)\\
\end{array}
\right)\,.
\ee
Define now
\be
\mathcal{A}^{r,s}_{n,m}(\lambda)=A_{n,m}\ast C_{r,s}(\lambda)=\delta_{r,s}A_{n,m}(\lambda)-\delta_{\bar{r},s}b_{n,m}(\lambda)\,,
\ee
where
\be
\bar{r}=2\delta_{r,1}+\delta_{r,2}\,,
\label{r_bar}
\ee
while $A_{n,m}$ and $b_{n,m}$ are defined in \eqref{eq:A_nm} and \eqref{eq:b_mn} respectively. One can straightforwardly verify the following identities
\bea
\mathcal{A}_{n,m}^{r,s}(\lambda)-s\ast \left(\mathcal{A}_{n-1,m}^{r,s}+\mathcal{A}_{n+1,m}^{r,s}\right)(\lambda)&=&\delta_{n,m}\delta_{r,s}\delta(\lambda)-\delta_{n,m}\delta_{r,\bar{s}}s(\lambda)\,,\quad n\geq 2\,,\label{eq:identity_kernel_1}\\
\mathcal{A}_{1,m}^{r,s}(\lambda)-s\ast\mathcal{A}_{2,m}^{r,s}(\lambda)&=&\delta_{1,m}\delta_{r,s}\delta(\lambda)-\delta_{1,m}\delta_{r,\bar{s}}s(\lambda)\,.
\label{eq:identity_kernel_2}
\eea
Plugging \eqref{eq:identity_kernel_1} into \eqref{eq:thermo_bethe_equations}  for $n\geq 2$, we obtain
\be
\rho^{(r)}_{h,n}(\lambda)+s\ast\sum_{m,s}\left(\mathcal{A}^{r,s}_{n-1,m}+\mathcal{A}^{r,s}_{n+1,m}\right)\ast\rho^{(s)}_m(\lambda)+\rho_n^{(r)}(\lambda)-s\ast \rho_{n}^{(\bar{r})}(\lambda)=a_{n}(\lambda)\delta_{r,1}\,.
\ee
Using again \eqref{eq:thermo_bethe_equations} to remove the infinite sum, we readily obtain
\be
\rho_{t,n}^{(r)}(\lambda)=s\ast \left[\rho^{(r)}_{h,n-1}+\rho^{(r)}_{h,n+1}\right](\lambda)+s\ast \rho_{n}^{(\bar{r})}(\lambda)\,,
\ee
where we used \eqref{eq:s-a_identity}. This is precisely \eqref{eq:tri-diagonal_BT1}. Analogously, using identity \eqref{eq:identity_kernel_2} for $n=1$, one has
\be
\rho^{(r)}_{h,1}(\lambda)+s\ast\sum_{m,s}\mathcal{A}^{r,s}_{2,m}\ast\rho^{(s)}_m(\lambda)+\rho_1^{(r)}(\lambda)-s\ast \rho_{1}^{(\bar{r})}(\lambda)=a_{1}(\lambda)\delta_{r,1}\,,
\ee
which, removing the infinite sum using \eqref{eq:thermo_bethe_equations} and using \eqref{eq:fourier_s}, is cast into
\be
\rho_{t,1}^{(r)}(\lambda)=s(\lambda)+s\ast \rho^{(r)}_{h,2}(\lambda)+s\ast \rho_{1}^{(\bar{r})}(\lambda)\,,
\ee
which coincides with \eqref{eq:tri-diagonal_BT2}.

\section{Partially decoupled form of the saddle-point equations}\label{sec:derivation_decoupled}

In this appendix we derive the partially decoupled form of the saddle-point equations \eqref{eq:eta_(1)}\,--\,\eqref{eq:eta_(2)}. 

Using the identities \eqref{eq:identity_kernel_1}\,--\,\eqref{eq:identity_kernel_2}, one can immediately derive from \eqref{eq:eta_(1)}\,--\,\eqref{eq:eta_(2)}
\bea
\ln\eta_1^{(r)}(\lambda)&=&d_1(\lambda)+s\ast \ln [1+\eta_2^{(r)}](\lambda)-s\ast \ln \left[1+\left(\eta_1^{(\bar{r})}\right)^{-1}\right](\lambda)\,,\label{decoupled1}\\
\ln\eta_n^{(r)}(\lambda)&=&d_n(\lambda)+s\ast\left( \ln [1+\eta_{n-1}^{(r)}]+\ln [1+\eta_{n+1}^{(r)}]\right)(\lambda)-s\ast \ln \left[1+\left(\eta_n^{(\bar{r})}\right)^{-1}\right](\lambda)\,,\label{decoupled2}
\eea
where we used the definition \eqref{r_bar}, while
\be
d_{n}(\lambda)=g_n(\lambda)-s\ast g_{n+1}(\lambda)-s\ast g_{n-1}(\lambda)\,,
\label{temporary_kernels}
\ee
where $g_n$ is given by \eqref{eq:g_function} and $g_0\equiv 0$. The kernels \eqref{temporary_kernels} can be simplified by systematic use of the convolution theorem. Using some known results on the Fourier transform of generalized functions~\cite{vladimirovbook}, we find the following identity  
\be
\mathcal{F}\left[\ln\left( x^2+\frac{a^2}{4}\right)\right](k)=\frac{2\pi\left(1-e^{-\frac{1}{2} |a k|}\right)}{|k|}-2\pi\mathcal P\frac{1}{|k|}-4\pi\gamma \delta(k)\,.
\label{eq:identity}
\ee
Here $\gamma$ is the Euler constant $\gamma\simeq 0.577$ and $\mathcal P[1/|x|]$ is such that
\be
\int {\rm d}x \,\, f(x) \,\,\mathcal P\left[\frac{1}{|x|}\right]\equiv\int\limits_{|x|<1} {\rm d}x \,\,\frac{f(x)-f(0)}{|x|}+\int\limits_{|x|>1} {\rm d}x \,\,\frac{f(x)}{|x|}\,,
\ee
with $f(x)$ smooth. 

The identity~\eqref{eq:identity} allows us to establish
\be
\left[s\ast \left(f_{n}+f_{n+2}\right)\right](\lambda)=f_{n+1}(\lambda)\,,\qquad n\geq 0\,,
\ee
where $f_n(\lambda)$ is defined in \eqref{eq:f_function}. Then, in the case of $n$ odd, we find 
\bea
g_n(\lambda)-s\ast \left(g_{n-1}+g_{n+1}\right)(\lambda)
&=&f_0(\lambda)-2s\ast f_1(\lambda)\,.
\eea
Analogously, if $n$ is even we have 
\bea
g_n(\lambda)-s\ast \left(g_{n-1}+g_{n+1}\right)(\lambda)
&=&-f_0(\lambda)+2s\ast f_1(\lambda)\,.
\eea
Finally, using again \eqref{eq:identity} we obtain  
\be
f_0(\lambda)-2s\ast f_1(\lambda)=-\int_{-\infty}^{\infty}{\rm d}k\,\, e^{-i\lambda k}\frac{\tanh(k/2)}{k}=\ln\left[\tanh^{2}\left(\frac{\pi \lambda}{2}\right)\right]\,.
\ee
This proves \eqref{eq:driving_dn}.

For completeness, we now show how to cast \eqref{decoupled1}\,--\,\eqref{decoupled2} in yet another form which is similar to the integral equations derived in the finite temperature case~\cite{jls-89}. The derivation closely follows Ref.~\cite{afl-83}.  First, we define
\bea
Q_{n}^{(r)}(\lambda)&=&\ln \left[1+\left(\eta_n^{(r)}\right)^{-1}\right](\lambda)\,,\\
h_{n}^{(r)}(\lambda)&=&s^{-1}\left[d_n\right](\lambda)-s^{-1}\left[\ln\left(1+\eta_n^{(r)}\right)\right](\lambda)+\ln(1+\eta_{n+1}^{(r)})(\lambda)+\ln(1+\eta_{n-1}^{(r)})(\lambda)\,,\label{eq:hfunction}
\eea
where $s^{-1}[\mathcal{G}](\lambda)$ is formally defined as
\be
s^{-1}[\mathcal{G}](x)=\frac{1}{2\pi}\int_{-\infty}^{\infty}\!\!{\rm d}x\, 2\cosh(k/2)\hat{\mathcal{G}}(k)e^{-ikx}\,,
\ee
namely it is the inversion of the integral functional with kernel $s$ defined in \eqref{eq:s_function}. In \eqref{eq:hfunction} we used the convention
\be
\eta^{(r)}_0(\lambda)\equiv 0 \,.
\label{convention_eta}
\ee
Equations \eqref{decoupled1}\,--\,\eqref{decoupled2} can then be rewritten as
\be
s^{-1}\left[Q_{n}^{(r)}\right](\lambda)=Q^{(\bar r)}_n(\lambda)-h^{(r)}_n(\lambda)\,,
\label{formal_equation}
\ee
where $\bar r$ is defined in \eqref{r_bar}. A formal solution of \eqref{formal_equation} can be obtained in Fourier transform. It is given by
\bea
\mathcal F\left[{Q}_{n}^{(1)}(\lambda)\right](k)&=&-\mathcal F\left[h_{n}^{(1)}(\lambda)\right](k)\frac{\sinh(k)}{\sinh(3k/2)}-\mathcal F\left[h_{n}^{(2)}(\lambda)\right](k)\frac{\sinh(k/2)}{\sinh(3k/2)}\,,\label{aux_fourier1}\\
\mathcal F\left[{Q}_{n}^{(2)}(\lambda)\right](k)&=&-\mathcal F\left[h_{n}^{(1)}(\lambda)\right](k)\frac{\sinh(k/2)}{\sinh(3k/2)}-\mathcal F\left[h_{n}^{(2)}(\lambda)\right](k)\frac{\sinh(k)}{\sinh(3k/2)}\,.\label{aux_fourier2}
\eea
Using the expression for the inverse Fourier transform \cite{gr-14}
\be
\mathcal{F}^{-1}\left[\frac{\sinh(\alpha k)}{\sinh(\beta k)}\right](x)=\frac{1}{2\beta}\frac{\sin(\pi\alpha/\beta)}{\left[\cosh(\pi x/\beta)+\cos(\pi \alpha/\beta)\right]}\,,
\ee
one can directly take the inverse Fourier transform of \eqref{aux_fourier1}\,--\,\eqref{aux_fourier2}. Defining
\bea
\mathcal{G}_{1}(x)&\equiv & \mathcal{F}^{-1}\left[\frac{\sinh(k/2)}{\sinh( 3k/2)}\right](x)= \frac{1}{\sqrt{3}}\frac{1}{\left(2\cosh(2\pi x/3)+1\right)}\,,\\
\mathcal{G}_{2}(x)&\equiv & \mathcal{F}^{-1}\left[\frac{\sinh(k)}{\sinh( 3k/2)}\right](x)= \frac{1}{\sqrt{3}}\frac{1}{\left(2\cosh(2\pi x/3)-1\right)}\,,
\eea
and using 
\bea
\frac{2\cosh(k/2)\sinh(k)}{\sinh(3k/2)}&=&\frac{\sinh(k/2)}{\sinh(3k/2)}+1\,,\\
\frac{2\cosh(k/2)\sinh(k/2)}{\sinh(3k/2)}&=&\frac{\sinh(k)}{\sinh(3k/2)}\,,
\eea
we obtain 
\bea
\ln \eta^{(1)}_n(\lambda) &=&w_n(\lambda)+\mathcal{G}_2\ast \ln\left[\frac{\left(1+\eta^{(1)}_{n+1}\right)\left(1+\eta^{(1)}_{n-1}\right)}{\left(1+\eta^{(2)}_{n}\right)}\right](\lambda) +\mathcal{G}_1\ast \ln\left[\frac{\left(1+\eta^{(2)}_{n+1}\right)\left(1+\eta^{(2)}_{n-1}\right)}{\left(1+\eta^{(1)}_{n}\right)}\right](\lambda)\,,\label{appeq:eta_decoupled1}\\
\ln \eta^{(2)}_n(\lambda) &=&w_n(\lambda)+\mathcal{G}_2\ast \ln\left[\frac{\left(1+\eta^{(2)}_{n+1}\right)\left(1+\eta^{(2)}_{n-1}\right)}{\left(1+\eta^{(1)}_{n}\right)}\right](\lambda) +\mathcal{G}_1\ast \ln\left[\frac{\left(1+\eta^{(1)}_{n+1}\right)\left(1+\eta^{(1)}_{n-1}\right)}{\left(1+\eta^{(2)}_{n}\right)}\right](\lambda) \,,\label{appeq:eta_decoupled2}
\eea
with the convention \eqref{convention_eta} and where the driving terms are given by
\be
w_n(\lambda)=\frac{1}{2\pi}\int_{-\infty}^{\infty}{\rm d}k\ \hat{d}_n(k)\frac{2\cosh(k/2)}{2\cosh(k/2)-1}e^{-ik \lambda}=(-1)^n\int_{-\infty}^{\infty}\frac{{\rm d}k}{k}\frac{\sinh(k/2)}{\left(\cosh(k/2)-1/2\right)}e^{-ik \lambda}\,.
\ee
This integral can be performed explicitly.  We use \cite{gr-14}
\be
\mathcal{F}\left[\frac{\cosh(ax)}{\sinh(bx)}\right](k)=i\pi\frac{\sinh(\pi k/b)}{b\left(\cosh(\pi k/b)+\cos(\pi a/b)\right)}\,,\qquad\qquad |a|<|b|\,.
\ee
Choosing $b=2\pi$, $a=4\pi/3$, and exploiting the properties of the Fourier transform, one arrives at
\be
-\frac{1}{2\pi}\int_{-\infty}^{\infty}\frac{{\rm d}k}{k}\frac{\sinh(k/2)}{\cosh(k/2)-1/2}e^{ikx}=\frac{1}{2\pi} \ln\left[\frac{-1 + 2 \cosh(2 \pi x/3)}{1 + 2 \cosh(2 \pi x/3)} \tanh^2(\pi x/3)\right]\,.
\ee
Note that \eqref{appeq:eta_decoupled1}\,--\,\eqref{appeq:eta_decoupled2} are exactly of the same form of the decoupled thermal equations reported in \cite{jls-89}. As expected from the experience with previous applications of the Quench Action, the only difference is in the driving terms $w_n(\lambda)$.

\section{Higher local conserved charges}\label{sec:derivation_charges}
In this appendix we discuss the computation of the initial-state expectation values of local conserved charges. More specifically, we detail the derivation of formula \eqref{eq:final_generating_function} for the corresponding generating function. 

As we anticipated in the main text, we exploit the method used in \cite{fe-13} for the case of the XXZ spin-$1/2$ chain. The starting point of our derivation is provided by the formal expansion of the transfer matrix \eqref{eq:transfer_matrix} in terms of the local charges \eqref{eq:charges_definition}, namely
\bea
\tau(\lambda)=\exp\left(-i\sum_{k=1}^{\infty}Q_{k+1}\frac{\lambda^{k}}{k!}\right)\,.
\label{eq:formal_expansion}
\eea
Since the the operators $Q_k$ are self-adjoint, Eq.~\eqref{eq:formal_expansion} suggests that for large $L$ one can write
\bea
\tau^{-1}(\lambda)=\tau^{\dagger}(\lambda)\,,
\label{eq:inversion_relation}
\eea
in the sense that the power series expansions in $\lambda$ of the two sides of \eqref{eq:inversion_relation} coincide for $\lambda\in\mathbb{R}$. Next, from \eqref{eq:def_generating_function} we observe that the generating function $\Omega_{\Psi_0}(\lambda)$ can be defined as 
\bea
-i \Omega_{\Psi_0}(\lambda)=\lim_{L\rightarrow\infty}\frac{1}{L}\langle \Psi_0|\frac{\partial}{\partial \mu}\log(\tau(\mu))|\Psi_0\rangle\Big|_{\mu=\lambda}=\lim_{L\rightarrow\infty}\frac{1}{L}\langle \Psi_0|\tau^{-1}(\lambda)\frac{\partial}{\partial \mu}\tau(\mu)|\Psi_0\rangle\Big|_{\mu=\lambda}\,.
\label{eq:def_generating_function2}
\eea
Using \eqref{eq:inversion_relation} we then obtain
\bea
\Omega_{\Psi_0}(\lambda)=i \lim_{L\rightarrow\infty}\frac{1}{L}\langle \Psi_0|\tau^{\dagger}(\lambda)\frac{\partial}{\partial \mu} \tau(\mu)|\Psi_0\rangle\Big|_{\mu=\lambda}\,.
\label{eq:final_expectation_value}
\eea
The computation is then reduced to evaluating the expectation value of the product of two transfer matrices. This can be performed analytically, due to the simple structure of our initial state and exploiting the representation of transfer matrix \eqref{eq:transfer_matrix} as a matrix product operator. In fact, the steps involved for the explicit evaluation of the r.h.s. of \eqref{eq:final_expectation_value} are completely analogous to those explained in \cite{fcec-14} for the case of the spin-$1/2$ chain and will not be reported here. Note in particular that the derivative in \eqref{eq:final_expectation_value} can be efficiently performed by means of the Jacobi formula \cite{fcec-14}, which makes it possible to reach a closed form analytical result. The intermediate analytical calculations can be easily carried out with the program Mathematica, so that one finally derives Eq.~\eqref{eq:final_generating_function}.

\section{Elementary excitations over a stationary state}
\label{app:velocities}

Here we consider excitations on a stationary state described by a set of root densities $\{\rho_n^{(1)}(k),\rho_n^{(2)}(\lambda)\}$. We consider in a large, finite volume $L$ and take the system in the stationary state described by the set of integers $\{I_\alpha^n,J_\beta^m\}$, corresponding to a set of string centers $\{k^n_\alpha,\lambda_\beta^m\}$ through Eqs.~\eqref{eq:logarithmicBTE}. Using the counting functions \eqref{eq:counting1}\,--\,\eqref{eq:counting2} we can establish a one to one correspondence between the integers $\{I_\alpha^n,J_\beta^m\}$ and the rapidities $\{k^n_\alpha,\lambda_\beta^m\}$. The integers $\{I_\alpha^n,J_\beta^m\}$ are chosen such that the distribution of string centers describing the state in the thermodynamic limit are $\{\rho_n^{(1)}(k),\rho_n^{(2)}(\lambda)\}$. 

Excitations on the state are constructed by changing the value of  a finite number of integers in $\{I_\alpha^n,J_\beta^m\}$. The corresponding modified set of rapidities can be described as follows. The rapidities corresponding to the unchanged integers $\{\tilde k^n_\alpha,\tilde\lambda_\beta^m\}$; the ones of particle excitations $\{k^{p}_n,\lambda_m^{p}\}$, corresponding to the added integers; those of hole excitations $\{k^{h}_n,\lambda_m^{h}\}$, corresponding to the removed integers. The latter set is composed by fictitious rapidities, cancelling the contribution of those corresponding to the removed integers. The rapidities $\{\tilde k^n_\alpha,\tilde\lambda_\beta^m\}$ in the state with $n^{(r)}_{p,m}$  particle excitations and $n^{(r)}_{h,m}$ hole excitations for every species $r=1,2$ and string type $m=1,\ldots$ are related to $\{ k^n_\alpha,\lambda_\beta^m\}$ as    
\bea
\tilde k^n_\alpha - k^n_\alpha &=& \sum_m\sum_{\beta=1}^{n^{(1)}_{{p},m}} \frac{F^{11}_{nm}(k^n_\alpha|k^{p\,m}_\beta)}{L\rho^{(1)}_{t,n}(k^n_\alpha)}-\sum_m\sum_{\beta=1}^{n^{(1)}_{{h},m}} \frac{F^{11}_{nm}(k^n_\alpha|k^{h\,m}_\beta)}{L\rho^{(1)}_{t,n}(k^n_\alpha)}\nonumber\\
&&+\sum_m\sum_{\beta=1}^{n^{(2)}_{{p},m}} \frac{F^{12}_{nm}(k^n_\alpha|\lambda^{p\,m}_\beta)}{L\rho^{(1)}_{t,n}(k^n_\alpha)}-\sum_m\sum_{\beta=1}^{n^{(2)}_{{h},m}} \frac{F^{12}_{nm}(k^n_\alpha|\lambda^{h\,m}_\beta)}{L\rho^{(1)}_{t,n}(k^n_\alpha)}\,,\label{eq:definition1}\\
\tilde \lambda^n_\alpha - \lambda^n_\alpha &=& \sum_m\sum_{\beta=1}^{n^{(1)}_{{p},m}} \frac{F^{21}_{nm}(\lambda^n_\alpha|k^{p\,m}_\beta)}{L\rho^{(2)}_{t,n}(\lambda^n_\alpha)}-\sum_m\sum_{\beta=1}^{n^{(1)}_{{h},m}} \frac{F^{21}_{nm}(\lambda^n_\alpha|k^{h\,m}_\beta)}{L\rho^{(2)}_{t,n}(\lambda^n_\alpha)}\nonumber\\
&&+\sum_m\sum_{\beta=1}^{n^{(2)}_{{p},m}} \frac{F^{22}_{nm}(\lambda^n_\alpha|\lambda^{p\,m}_\beta)}{L\rho^{(2)}_{t,n}(\lambda^n_\alpha)}-\sum_m\sum_{\beta=1}^{n^{(2)}_{{h},m}} \frac{F^{22}_{nm}(\lambda^n_\alpha|\lambda^{h\,m}_\beta)}{L\rho^{(2)}_{t,n}(\lambda^n_\alpha)}\,,\label{eq:definition2}
\eea
where we introduced the \emph{shift functions} $F^{11}_{nm}(x|y), F^{12}_{nm}(x|y), F^{21}_{nm}(x|y)$, $F^{22}_{nm}(x|y)$ and assumed the total number of excitations to be even. The shift functions are defined through the following integral equations 
\bea
F^{11}_{nm}(x|y) &=& \frac{1}{2\pi}\Xi_{n,m}(x-y) -\sum_k\!\!\int\!{\rm d}z\,\,a_{n,k}(x-z)\vartheta^{(1)}_k(z)F^{11}_{km}(z|y)\notag\\
&&\qquad\qquad\qquad\quad+\sum_k\!\!\int\!{\rm d}z\,\,b_{n,k}(x-z)\vartheta^{(2)}_k(z)F^{21}_{km}(z|y)\,,\label{eq:F11}\\
F^{12}_{nm}(x|y) &=& -\frac{1}{2\pi}\Theta_{n,m}(x-y)-\sum_k\!\!\int\!{\rm d}z\,\,a_{n,k}(x-z)\vartheta^{(1)}_k(z)F^{12}_{km}(z|y)\notag\\
&&\qquad\qquad\qquad\qquad+\sum_k\!\!\int\!{\rm d}z\,\,b_{n,k}(x-z)\vartheta^{(2)}_k(z)F^{22}_{km}(z|y)\,,\label{eq:F12}\\
F^{21}_{nm}(x|y) &=& -\frac{1}{2\pi}\Theta_{n,m}(x-y)-\sum_k\!\!\int\!{\rm d}z\,\,a_{n,k}(x-z)\vartheta^{(2)}_k(z)F^{21}_{km}(z|y)\notag\\
&&\qquad\qquad\qquad\qquad+\sum_k\!\!\int\!{\rm d}z\,\,b_{n,k}(x-z)\vartheta^{(1)}_k(z)F^{11}_{km}(z|y)\,,\label{eq:F21}\\
F^{22}_{nm}(x|y) &=& \frac{1}{2\pi}\Xi_{n,m}(x-y) -\sum_k\!\!\int\!{\rm d}z\,\,a_{n,k}(x-z)\vartheta^{(2)}_k(z)F^{22}_{km}(z|y)\notag\\
&&\qquad\qquad\qquad\quad+\sum_k\!\!\int\!{\rm d}z\,\,b_{n,k}(x-z)\vartheta^{(1)}_k(z)F^{12}_{km}(z|y)\,,\label{eq:F22}
\eea
where the terms $\Xi_{n,m}$ and $\Theta_{n,m}$ are defined in \eqref{eq:Xi_mn} and \eqref{eq:Theta_mn}, while the kernels $a_{n,m}$ and $b_{n,m}$ are given in \eqref{eq:a_mn} and \eqref{eq:b_mn}. These equations are derived following the same procedure employed in the study of excitations on thermal states \cite{kbi-93}. Using the Bethe equations for the state described by $\{I_\alpha^n,J_\beta^m\}$ and those for the state with the excitations one finds 
\be
z^{(1)}_{\rm ex}(\tilde k^{n}_\alpha)-z^{(1)}(k^{n}_\alpha)=0\qquad\qquad z^{(2)}_{\rm ex}(\tilde \lambda^{n}_\alpha)-z^{(2)}(\lambda^{n}_\alpha)=0\,.\label{eq:difference}
\ee
Here $z^{(a)}(x)$ are the counting functions of the state, while $z_{\rm ex}^{(a)}(x)$ are those of the state with the excitations. Taking the thermodynamic limit of \eqref{eq:difference} and using the definition \eqref{eq:definition1}, \eqref{eq:definition2} one readily finds that the shift functions satisfy Eqs.~\eqref{eq:F11}\,--\,\eqref{eq:F22}.

One can easily compute the expectation value on excited states of the local conserved charges discussed in appendix~\ref{sec:derivation_charges}. Using \eqref{eq:finitesizeQ}, one can introduce for a given local conserved operator $Q$ the corresponding bare charge densities $q_n(\lambda)$ by
\be
\braket{\{I_\alpha^n,J_\beta^m\}|  Q|\{I_\alpha^n,J_\beta^m\}}=\sum_{n=1}^{\infty}\sum_{\alpha=1}^{M^{(1)}_n} q_n( k^n_\alpha)\,.
\ee
Then, the difference in the expectation value of the excited state and the one without excitations is given by  
\bea
\Delta Q&=&\sum_{n=1}^{\infty}\sum_{\alpha=1}^{M^{(1)}_n} \left(q_n(\tilde k^n_\alpha)-q_n(k^n_\alpha)\right)+\sum_{n=1}^{\infty}\sum_{\beta=1}^{n^{(1)}_{{p},n}}  q_n(k^{p\,n}_\beta)-\sum_{n=1}^{\infty}\sum_{\beta=1}^{n^{(1)}_{{h},n}}  q_n(k^{h\,n}_\beta)\notag\\
&=&\sum_{n=1}^{\infty}\sum_{\beta=1}^{n^{(1)}_{{p},n}}  q^{d\,(1)}_n(k^{p\,n}_\beta)-\sum_{n=1}^{\infty}\sum_{\beta=1}^{n^{(1)}_{{h},n}}  q^{d\,(1)}_n(k^{h\,n}_\beta)+\sum_{n=1}^{\infty}\sum_{\beta=1}^{n^{(2)}_{{p},n}}  q^{d\,(2)}_n(\lambda^{p\,n}_\beta)-\sum_{n=1}^{\infty}\sum_{\beta=1}^{n^{(2)}_{{h},n}}  q^{d\,(2)}_n(\lambda^{h\,n}_\beta)\,.
\eea
Here we introduced the \emph{dressed charges}
\bea
q^{d\,(1)}_n(x) &=& q_n(x)+\sum_k\!\!\int\!{\rm d}z\,\,q^{\prime}_{k}(z)\vartheta^{(1)}_k(z)F^{11}_{kn}(z|x)\,,\\
q^{d\,(2)}_n(x) &=& \sum_k\!\!\int\!{\rm d}z\,\,q^{\prime}_{k}(z)\vartheta^{(1)}_k(z)F^{12}_{kn}(z|x)\,,
\eea
where $q^{\prime}_{k}(z)=({\rm d}/{{\rm d}z})q_{k}(z)$. In particular, dressed energy and momentum are given by 
\bea
\varepsilon^{d\,(a)}_n(x) &=& \varepsilon_n(x)\delta_{a,1}+\sum_k\!\!\int\!{\rm d}z\,\,\varepsilon^{\prime}_{k}(z)\vartheta^{(1)}_k(z)F^{1a}_{kn}(z|x)\,,\qquad\qquad\qquad a=1,2\,,\\
p^{d\,(a)}_n(x) &=& p_n(x)\delta_{a,1}+\sum_k\!\!\int\!{\rm d}z\,\,p^{\prime}_{k}(z)\vartheta^{(1)}_k(z)F^{1a}_{kn}(z|x)\,,\qquad\qquad\qquad a=1,2\,,
\eea
where $p_n(\lambda)$ and $\varepsilon_n(\lambda)$ are given in \eqref{eq:pn} and \eqref{epsilon_energy} respectively. We are now in a position to formally solve the equations \eqref{eq:F11}--\eqref{eq:F22}, finding a closed expression for the dressed charges. In doing so, it is convenient to introduce the following compact notations~\cite{BCDF:transport}. We represent functions $w_{j}(\lambda)$ 
by a vector $\vec w$, such that
\be
\label{eq:vec}
[ \vec w ]_{j}(\lambda)= w_{j}(\lambda)\,.
\ee
Kernels $A_{jk}(\lambda,\mu)$ which are functions of two rapidities with two string indices are instead represented by an operator $\hat{A}$ acting as follows
\be
[{\hat{A}} \vec w ]_{j}(\lambda)=\sum\nolimits_{k}\int{{\rm d}\mu}\, A_{jk}(\lambda,\mu) w_k(\mu)\,.
\label{eq:def_matrix_op}
\ee
The inverse of  $\hat{A}$ is the operator $\hat A^{-1}$ satisfying
\be
\sum\nolimits_{k}\int{{\rm d}\nu}\, [A^{-1}]_{i k}(\lambda,\nu)A_{kj}(\nu,\mu)=\delta(\lambda-\mu)\delta_{i j}\, .
\ee
In this way, the distribution $\delta(\lambda-\mu)\delta_{i j}$ corresponds to the identity $\hat 1$. It is also convenient to define \emph{diagonal} operators $\hat w$ associated with a function of a single rapidity
\be
[\hat w]_{i j}(\lambda,\mu)=\delta(\lambda-\mu)\delta_{i j}w_i(\lambda)\, .
\ee

Rewriting \eqref{eq:F11}--\eqref{eq:F22} in compact notations we have 
\bea
\hat F^{11} &=&\frac{1}{2\pi}\hat \Xi  -\hat a \hat \vartheta^{(1)}\hat F^{11} + \hat b \hat\vartheta^{(2)} \hat F^{21}\,,\label{eq:F11c}\\
\hat F^{12} &=&-\frac{1}{2\pi}\hat \Theta -\hat a \hat\vartheta^{(1)} \hat F^{12}+\hat b\hat \vartheta^{(2)}\hat F^{22}\,,\label{eq:F12c}\\
\hat F^{21} &=& - \frac{1}{2\pi}\hat \Theta  -\hat a \hat \vartheta^{(2)}\hat F^{21} + \hat b \hat\vartheta^{(1)} \hat F^{11}\,,\label{eq:F21c}\\
\hat F^{22} &=&\frac{1}{2\pi}\hat \Xi-\hat a \hat\vartheta^{(2)} \hat F^{22}+\hat b\hat \vartheta^{(1)}\hat F^{12}\,,\label{eq:F22c}
\eea
where $\hat \vartheta^{(r)}$ are diagonal operators
\be
[\hat \vartheta^{(c)}]_{ij}(\lambda,\mu)=\delta(\lambda-\mu)\delta_{ij} \vartheta^{(c)}(\lambda)\,,\qquad c=1,2\,,
\ee
while $\hat a$, $\hat b$ and $\hat F^{rs}$ are non diagonal 
\be
[\hat a]_{ij}(\lambda,\mu)=a_{i,j}(\lambda-\mu)\,,\qquad [\hat b]_{ij}(\lambda,\mu)=b_{i,j}(\lambda-\mu)\,,\qquad [\hat F^{rs}]_{ij}(\lambda,\mu)=F^{rs}_{ij}(\lambda|\mu)\,,\qquad r,s=1,2\,.
\ee
Equations \eqref{eq:F11c}\,-\,\eqref{eq:F22c} are coupled in pairs. Let us consider the first pair composed by \eqref{eq:F11c} and \eqref{eq:F21c}, from the latter we find 
\be
\hat F^{21} = -\frac{1}{2\pi} (\hat 1 + \hat a \hat \vartheta^{(2)})^{-1}\hat\Theta+(\hat 1 + \hat a \hat \vartheta^{(2)})^{-1}\hat b \hat \vartheta^{(1)} \hat F^{11}\,. 
\ee
Plugging into \eqref{eq:F11c} we have 
\be
\left(\hat 1 +\hat a \hat \vartheta^{(1)}- \hat b \hat\vartheta^{(2)}(\hat 1 + \hat a \hat \vartheta^{(2)})^{-1}\hat b \hat \vartheta^{(1)} \right)\hat F^{11} = \frac{1}{2\pi}\hat\Xi-\frac{1}{2\pi}\left(\hat b \hat\vartheta^{(2)} (\hat 1 + \hat a \hat \vartheta^{(2)})^{-1}\right)\hat\Theta\,,
\ee
so that
\bea
\hat F^{11} &=& \frac{1}{2\pi}\left(\hat 1 +\hat a \hat \vartheta^{(1)}- \hat b \hat\vartheta^{(2)}(\hat 1 + \hat a \hat \vartheta^{(2)})^{-1}\hat b \hat \vartheta^{(1)} \right)^{-1}\left(\hat\Xi-\left(\hat b \hat\vartheta^{(2)} (\hat 1 + \hat a \hat \vartheta^{(2)})^{-1}\right)\hat\Theta\right) \,.\label{eq:F11explicit}
\eea
Analogously, considering \eqref{eq:F12c} and \eqref{eq:F22c} we have
\bea
\hat F^{12} &=& \frac{1}{2\pi}\left(\hat 1 +\hat a \hat \vartheta^{(1)}- \hat b \hat\vartheta^{(2)}(\hat 1 + \hat a \hat \vartheta^{(2)})^{-1}\hat b \hat \vartheta^{(1)} \right)^{-1}\left(\left(\hat b \hat\vartheta^{(2)} (\hat 1 + \hat a \hat \vartheta^{(2)})^{-1}\right)\hat\Xi-\hat\Theta\right) \,.\label{eq:F11explicit2}
\eea
In compact notation, the equations for the dressed charges read as 
\be
\vec q^{\,\,d\,(a)} = \vec q \,\,\delta_{a,1}+( \hat F^{1a})^t\hat\vartheta^{(1)}\vec q^{\,\,\prime} \qquad\qquad\qquad a=1,2\,\,,\label{eq:dressedchargesc}
\ee
where $t$ is the transposition operation. For a given kernel $A$, acting as in \eqref{eq:def_matrix_op}, the transposition is defined by 
\be
\left[A^{t}\vec w\right]_j(\lambda)=\sum_k\int {\rm d}\mu \, A_{kj}(\mu,\lambda)w_k(\mu)\,.
\ee
Let us consider $a=1$ and plug \eqref{eq:F11explicit} into \eqref{eq:dressedchargesc} 
\bea
\vec q^{\,\,d\,(1)} &=& \vec q -\frac{1}{2\pi}  \left(\hat\Xi-\hat\Theta\left( (\hat 1 + \hat \vartheta^{(2)} \hat a)^{-1}\hat\vartheta^{(2)} \hat b \right)\right) \left(\hat 1 + \hat \vartheta^{(1)}\hat a-  \hat\vartheta^{(1)}\hat b(\hat 1 + \hat \vartheta^{(2)} \hat a)^{-1} \hat \vartheta^{(2)}\hat b \right)^{-1} \hat\vartheta^{(1)}\vec q^{\,\,\prime}\notag\\
&=& \vec q -\frac{1}{2\pi} \left(\hat\Xi \hat\vartheta^{(1)} -\hat\Theta\left( \hat\vartheta^{(2)}  (\hat 1 + \hat a \hat\vartheta^{(2)} )^{-1}\hat b \hat\vartheta^{(1)}  \right)\right)\left(\hat 1 + \hat a \hat\vartheta^{(1)}-  \hat b \hat \vartheta^{(2)} (\hat 1 + \hat a \hat \vartheta^{(2)})^{-1}\hat b\hat\vartheta^{(1)} \right)^{-1} \vec q^{\,\,\prime}\,,
\eea
where in the first step we used that $\hat\Theta$ and $\hat\Xi$ are antisymmetric, $\hat a$ and $\hat b$ are symmetric and $\hat \vartheta^{(r)}$ diagonal. Let us now consider the equation ``in components'' $[\cdot]_j(\lambda)$ and take the derivative with respect to $\lambda$. Re-expressing then everything in compact notations we have 
\be
\vec q^{\,\,d\,(1)\,\prime} = \left(\hat 1 + \hat a \hat\vartheta^{(1)}-  \hat b \hat \vartheta^{(2)} (\hat 1 + \hat a \hat \vartheta^{(2)})^{-1}\hat b\hat\vartheta^{(1)} \right)^{-1} \vec q^{\,\,\prime}\,,
\ee
where we used 
\bea
\frac{1}{2\pi}\partial_{\lambda}\Xi_{m,n}(\lambda,\mu)=a_{n,m}(\lambda,\mu)\,,\\
\frac{1}{2\pi}\partial_{\lambda}\Theta_{m,n}(\lambda,\mu)=b_{n,m}(\lambda,\mu)\,.
\eea
Analogously we find
\bea
\vec q^{\,\,d\,(2)\,\prime} &=&\left(\hat b \hat\vartheta^{(1)} -\hat a \hat\vartheta^{(2)}  (\hat 1 + \hat a \hat\vartheta^{(2)} )^{-1}\hat b \hat\vartheta^{(1)}  \right)\left(\hat 1 + \hat a \hat\vartheta^{(1)}-  \hat b \hat \vartheta^{(2)} (\hat 1 + \hat a \hat \vartheta^{(2)})^{-1}\hat b\hat\vartheta^{(1)} \right)^{-1} \vec q^{\,\,\prime}\notag\\
&=& (\hat 1 + \hat a \hat\vartheta^{(2)} )^{-1}\hat b \hat\vartheta^{(1)} \left(\hat 1 + \hat a \hat\vartheta^{(1)}-  \hat b \hat \vartheta^{(2)} (\hat 1 + \hat a \hat \vartheta^{(2)})^{-1}\hat b\hat\vartheta^{(1)} \right)^{-1} \vec q^{\,\,\prime}\,.
\eea
Applying our general results to the case of dressed momenta we have 
\bea
\vec p^{\,\,d\,(1)\,\prime} &=& \left(\hat 1 + \hat a \hat\vartheta^{(1)}-  \hat b \hat \vartheta^{(2)} (\hat 1 + \hat a \hat \vartheta^{(2)})^{-1}\hat b\hat\vartheta^{(1)} \right)^{-1} 2\pi \vec a=2 \pi\vec \rho_{t}^{\,\,(1)}\,,\\
\vec p^{\,\,d\,(2)\,\prime} &=& (\hat 1 + \hat a \hat\vartheta^{(2)} )^{-1}\hat b \hat\vartheta^{(1)} \left(\hat 1 + \hat a \hat\vartheta^{(1)}-  \hat b \hat \vartheta^{(2)} (\hat 1 + \hat a \hat \vartheta^{(2)})^{-1}\hat b\hat\vartheta^{(1)} \right)^{-1} 2\pi\vec a=2\pi \vec \rho_{t}^{\,\,(2)}
\eea    
where the second equality follows by inverting Eqs.~\eqref{eq:TBAexplicit1} -- \eqref{eq:TBAexplicit2}. We can finally write an integral equation for the \emph{group velocities} of the quasi-particle excitations. These are defined as~\cite{BCDF:transport, BEL:lightcone} 
\be
v^{(r)}_n(\lambda)\equiv\frac{{\rm d} \varepsilon^{d\,(r)}_n(\lambda)}{{\rm d} p^{d\,(r)}_n(\lambda)}=\frac{\varepsilon^{d\,(r)\,\prime}_n(\lambda)}{p^{d\,(r)\,\prime}_n(\lambda)}=\frac{\varepsilon^{d\,(r)\,\prime}_n(\lambda)}{2\pi \rho^{(r)}_{t,n}(\lambda)}\,.
\ee 
In compact notations we then have
\bea
\hat v^{(1)} \vec\rho_t^{\,\,(1)} &=& \frac{1}{2\pi} \vec\varepsilon^{\,\,d\,(1)\,\prime}=\frac{1}{2\pi} \left(\hat1+ \hat a \hat\vartheta^{(1)} -  \hat b \hat \vartheta^{(2)} (\hat 1 + \hat a \hat \vartheta^{(2)})^{-1}\hat b \hat\vartheta^{(1)} \right)^{-1} \vec \varepsilon^{\,\,\prime}\\
\hat v^{(2)} \vec\rho_t^{\,\,(2)} &=& \frac{1}{2\pi} \vec\varepsilon^{\,\,d\,(2)\,\prime}= \frac{1}{2\pi} (\hat1+ \hat a  \hat\vartheta^{(2)} )^{-1} \left((\hat\vartheta^{(1)})^{-1}\hat b^{-1} + \hat a \hat b^{-1}-  \hat b \hat \vartheta^{(2)} (\hat 1 + \hat a \hat \vartheta^{(2)})^{-1}\right)^{-1} \vec \varepsilon^{\,\,\prime}\,.
\eea
Note now that these equations have the exact same form of those obtained by inverting Eqs.~\eqref{eq:TBAexplicit1} -- \eqref{eq:TBAexplicit2}, the driving term $a$ being replaced by $\varepsilon^{\prime}/(2\pi)$. Then, it is straightforward to bring these equations to the following final explicit form
\bea
\rho^{(2)}_{t,n}(\lambda) v^{(2)}_n(\lambda)&=&\sum_{k}\left( b_{n,k}\ast v^{(1)}_k \rho^{(1)}_k\right)(\lambda)-\sum_{k} \left(a_{n,k}\ast v^{(2)}_k \rho^{(2)}_k\right)(\lambda)\,,\\
\rho^{(1)}_{t,n}(\lambda) v^{(1)}_n(\lambda)&=&\frac{1}{2\pi}\varepsilon^{\prime}_n(\lambda)-\sum_k \left(a_{n,k}\ast  \rho_k^{(1)} v^{(1)}_k\right)(\lambda)+\sum_k \left(b_{n,k}\ast \rho_k^{(2)} v^{(2)}_k\right)(\lambda)\,.
\eea

\end{document}